\documentclass[secnumarabic, graphics,floatfix, nofootinbib,tightenlines,nobibnotes, aps, prl, 12pt]{revtex4-1}
\usepackage{graphicx}
\usepackage[english]{babel}
\usepackage[utf8]{inputenc}
\usepackage{amsmath,amssymb}
\usepackage{amsfonts}
\usepackage{multirow}
\usepackage{pstricks}
\usepackage{float}
\usepackage{subfigure}
\usepackage{color}
\usepackage{epsfig}
\usepackage{pst-tools}
\begin{document}

\title{Dirac quasinormal modes of power-Maxwell charged black holes in Rastall gravity}

\author{Cai-Ying Shao$^{1}$}
\author{Yu Hu$^{1}$}
\author{Yu-Jie Tan$^{1}$\footnote{yjtan@hust.edu.cn}}
\author{Cheng-Gang Shao$^{1}$\footnote{cgshao@hust.edu.cn}}
\author{Kai Lin$^{2,3}$\footnote{lk314159@hotmail.com}}
\author{Wei-Liang Qian$^{3,4,5}$\footnote{wlqian@usp.br}}

\affiliation{$^{1}$ MOE Key Laboratory of Fundamental Physical Quantities Measurement, Hubei Key Laboratory of Gravitation and Quantum Physics, PGMF, and School of Physics, Huazhong University of Science and Technology, 430074, Wuhan, Hubei, China}
\affiliation{$^{2}$ Hubei Subsurface Multi-scale Imaging Key Laboratory, Institute of Geophysics and Geomatics, China University of Geosciences, 430074, Wuhan, Hubei, China}
\affiliation{$^{3}$ Escola de Engenharia de Lorena, Universidade de S\~ao Paulo, 12602-810, Lorena, SP, Brazil}
\affiliation{$^{4}$ Faculdade de Engenharia de Guaratinguet\'a, Universidade Estadual Paulista, 12516-410, Guaratinguet\'a, SP, Brazil}
\affiliation{$^{5}$ Center for Gravitation and Cosmology, School of Physical Science and Technology, Yangzhou University, 225002, Yangzhou, Jiangsu, China}

\date{April 28, 2020}

\begin{abstract}
In this paper, we study the quasinormal modes of the massless Dirac field for charged black holes in Rastall gravity.
The spherically symmetric black hole solutions in question are characterized by the presence of a power-Maxwell field, surrounded by the quintessence fluid.
The calculations are carried out by employing the WKB approximations up to the thirteenth order, as well as the matrix method.
The temporal evolution of the quasinormal modes is investigated by using the finite difference method.
Through numerical simulations, the properties of the quasinormal frequencies are analyzed, including those for the extremal black holes.
Among others, we explore the case of a second type of extremal black holes regarding the Nariai solution, where the cosmical and event horizon coincide.
The results obtained by the WKB approaches are found to be mostly consistent with those by the matrix method.
It is observed that the magnitudes of both real and imaginary parts of the quasinormal frequencies increase with increasing $\kappa $, the spin-orbit quantum number.
Also, the roles of the parameters $Q$ and $w$, associated with the electric charge and the equation of state of the quintessence field, respectively, are investigated regarding their effects on the quasinormal frequencies.
The magnitude of the electric charge is found to sensitively affect the time scale of the first stage of quasinormal oscillations, after which the temporal oscillations become stabilized.
It is demonstrated that the black hole solutions for Rastall gravity in asymptotically flat spacetimes are equivalent to those in Einstein gravity, featured by different asymptotical spacetime properties.
As one of its possible consequences, we also investigate the behavior of the late-time tails of quasinormal models in the present model.
It is found that the asymptotical behavior of the late-time tails of quasinormal modes in Rastall theory is governed by the asymptotical properties of the spacetimes of their counterparts in Einstein gravity.
\end{abstract}

\maketitle

\section{I. Introduction}\label{section1}
The energy-momentum conservation is a vital assumption in the derivation of the Einstein field equations.
As a fundamental concept, it has been soundly established for theoretical physics in Minkowski space.
However, to a certain degree, it becomes questionable and deserves further investigation in the context of the curved metric.
In 1972, the very question was raised by Peter Rastall~\cite{prd-rastall-01}, and subsequently, a modified gravitational theory was proposed, known as Rastall Gravity.
According to Rastall's prescription, the energy-momentum tensor conservation is modified to read
\begin{equation}\label{N1}
{T^v}_{\mu ;v} = \lambda {R_{,\mu }} .
\end{equation}
Here $R$ is the Ricci scalar, and $\lambda$ is a constant to be determined, which ensures the energy-momentum tensor conservation is restored when $\lambda = 0$.
As a matter of fact, Rastall gravity can be readily viewed~\cite{arxiv-rastall-mach-02} as an implementation that follows from Mach's principle.
Indeed, a ``Machian'' theory requires that the inertial of a massive local object is associated with the external energy-momentum distributions.
According to Rastall's recipe, the energy-momentum tensor, presented in the field equation as the source term, depends on the metric.
The latter is determined nonlocally, to be specific, by the energy momentum distribution of the entire spacetime.
Also, as the theory is obliged to agree with any existing measurement, in other words, its deviation from Einstein's theory of general relativity is subject to the astrophysical observations.
In particular, as the tests of general relativity on the largest scales is less stringent in comparison to those in the Solar System, it is plausible that intriguing results might be associated with large-scale phenomena, such as the evolution of the Universe, stellar structure, as well as gravitational collapse.

In fact, considerable efforts have been devoted to the researches concerning Rastall gravity.
The theory has been widely used in cosmology and black holes~\cite{jp-rastall-cosm-03,plb-rastall-Gravi-Potent-04,epjc-rastall-bd-05,ijmpd-rastall-bd-gde-06,grg-Thermodynamic-rastall}.
As it provides a new perspective for the study of gravitational systems, theoretical analyses have been carried out in the applications to galaxies regarding constraint in the observations~\cite{mnras-rastall-parame-gravilens-07}.
Besides, there are a variety of metric solutions about neutron stars~\cite{prd-rastall-Neut-Star-08}, wormholes~\cite{cjp-rastall-wormho-09}, and black holes~\cite{ahep-rastall-schw-ds-10,plb-blackhole-solution-rastall,Heydarzade:2016zof} inclusively in comparison with the corresponding solutions in Einstein's gravity.
In particular, the high-dimensional charged black hole metric surrounded by quintessence fluid has also been generalized to Rastall gravity recently~\cite{grg-rastall-hig-char-bh-11}.
The black hole solutions with the presence of both linear and nonlinear electromagnetic fields are obtained analytically, where the free parameters of the solution are further constrained by black hole thermodynamics.

As discussed above, Rastall's theory is potentially engaging in applications concerning gravitational collapse.
Therefore, in the present study, we focus on a specific problem concerning black hole quasinormal modes.
The latter is physically significant, mainly because they are closely related to the ringdown phase of the observed coalescence of binary systems.
Moreover, the topic is closely related to the no-hair theorem of black holes.
As an explicit test of the black hole perturbation theory, it might even be employed to analyze the stability of naked singularities~\cite{Chirenti:2014wza}.
Last but not least, the study of perturbations in the black hole background is essential for the AdS/CFT correspondence~\cite{ijmpa-qnm-ads/cft}.
By definition, the quasinormal mode oscillations essentially give an account of the temporal evolution triggered by initial perturbations in the black hole metric.
The resulting frequencies, known as quasinormal frequencies, are closely associated with intrinsic properties of the related non-perturbed spherical black holes.
The evolution of the black hole quasinormal modes can usually be divided into the following three stages~\cite{agr-qnm-review-03, agr-qnm-review-04}.
The irregular initial burst of oscillations characterizes the first stage.
Here, the frequency is mostly determined by the specific form of the initial disturbance.
The second stage is dominated by the quasinormal ringing, during which the oscillations become stabilized and do not depend on the initial conditions.
This stage is of particular physical interest since it only depends on the ``hairs'' of the black hole, such as mass, angular momentum, and electric charge.
Once experimentally observed, the extracted information can be utilized to provide strong evidence for the direct detection of the black hole.
As it is well-known, the distinctive exponential decay of quasinormal oscillation implies that the corresponding frequency is complex.
Its real part represents the magnitude of the oscillation, while the imaginary part indicates the decay rate.
The third and last stage is the late-time tail, which depends largely on the asymptotical form of the effective potential at spatial infinity.
For asymptotically flat spacetimes, it is due to the backscattering of perturbed wave packets by spacetime far away from the black hole~\cite{Ching:1995tj, Koyama:2000hj, Yu:2002st, Poisson:2002jz, Cardoso:2003jf, Jing:2004xv, Chen:2005vq}.
In this case, the disturbance is transformed from exponential decay to power-law decay, usually referred to as the power-law tail.
On the other hand, in asymptotically de Sitter spacetimes, the power-law behavior gives way to a more rapid, exponential decay~\cite{Brady:1996za, Brady:1999wd}.
Moreover, for the case of massive fields, the form of the late-time tail is found to be more complicated as it decays as well as oscillates~\cite{Koyama:2000hj, Koyama:2001ee, Jing:2004zb, Konoplya:2006gq}.

The study of quasinormal modes~\cite{agr-qnm-review-01, agr-qnm-review-03, agr-qnm-review-06, agr-qnm-review-02, agr-qnm-review-04, agr-qnm-review-06, Leaver:1986gd} was pioneered by the calculation of the gravitational perturbations in the Schwarzschild black hole~\cite{pr-qnm-stab-schw-01}.
Subsequently, it has been extended to various types of perturbations such as scalar fields~\cite{prd-qnm-scal-02,prd-scalar-qnm}, electromagnetic fields~\cite{grg-qnm-Ele-03,prd-electromagnetic-BTZ-qnm}, Dirac spinor~\cite{prd-qnm-dirac-04,prd-qnm-dirac-05} besides the gravitational ones~\cite{cqg-qnm-gravi-06}.
In addition to a few analytical solutions, the most important approaches for the calculation of quasinormal frequencies are semi-analytical or numerical schemes.
Among others, notable methods include the WKB approximation~\cite{aj-nm-semianal-wkb-01,prd-nm-wkb-02,prd-nm-wkb-03,Konoplya:2003ii,Matyjasek:2017psv,Konoplya:2019hlu,Matyjasek:2019eeu}, Posh-Teller potential approximation~\cite{pla-qnm-schw-01,prd-qnm-anal-02}, continuous fraction method~\cite{prs-qnm-anal-conti-01}, finite difference method~\cite{prd-qnm-lateti-Linear-01,prd-qnm-lateti-schw-02}, and Horowitz and Hubeny method~\cite{prd-qnm-ads-hh-01}.
In particular, the WKB method was generalized to the third order by Konoplya~\cite{Konoplya:2003ii}.
Recently, it was further extended to the thirteenth order by Matyjasek {\it et al. }~\cite{Matyjasek:2017psv, Matyjasek:2019eeu}, and its numerical implementation was subsequently developed and released publically by Konoplya {\it et al. }~\cite{Konoplya:2019hlu}.
The matrix method was also proposed recently~\cite{agr-qnm-lq-matrix-01,agr-qnm-lq-matrix-02} which has been shown to be competent approach for various different scenarios~\cite{agr-qnm-lq-matrix-03,agr-qnm-lq-matrix-04}.
It can be straightforwardly applied to various asymptotical spacetimes, and its accuracy can be improved as the number of interpolation points increases.

In the context of Rastall gravity, the quasinormal modes have been investigated by many authors~\cite{ctp-qnm-rastall,epjc-rastall-sclar-hig-quinte-13,gravitationalQNM-rastall,ahep-scalarqnm-rastall}.
To be specific, the effects of the Rastall parameter on the gravitational, electromagnetic, and massless scalar perturbations were explored in Ref.~\cite{ctp-qnm-rastall}.
The impacts of the spatial dimensions, equation of state of the surrounded quintessence field have been investigated in Ref.~\cite{epjc-rastall-sclar-hig-quinte-13}.
Moreover, the string and Rastall parameters were also found to affect the quasinormal frequencies in Ref.~\cite{gravitationalQNM-rastall}.
The present study, on the other hand, is devoted to investigating the quasinormal frequencies of the Dirac perturbations in the four-dimensional Rastall gravity.
The black hole metric in question is surrounded by the quintessence field, with the presence of the linear and nonlinear electromagnetic field.
We employed both the WKB method up to the thirteenth order as well as the matrix method to evaluate the quasinormal frequencies.
Also, the temporal evolution of smaller perturbations of the Dirac field is studied by using the finite difference method.
Besides, the present work is also focused on a distinctive feature regarding the late-time tails of the quasinormal modes.
As discussed in Refs.~\cite{Heydarzade:2016zof, Lin:2019gua}, the black hole solutions for Rastall gravity in asymptotically flat spacetimes might be equivalent to those in the de Sitter spacetimes.
Moreover, regarding the distinct characteristics for late-time decay in asymptotically flat and de Sitter spacetimes, it is interesting to explore the properties of the late-time tails in Rastall theory where the original spacetime is asymptotically flat.
Therefore, the present study is mostly concentrated on the second and third stages of quasinormal modes.

The paper is organized as follows.
In the next section, we briefly review the black hole solution obtained in Ref.~\cite{grg-rastall-hig-char-bh-11}.
In Section III, the master equation for the massless Dirac field is derived for spherically symmetric perturbations.
Section IV is devoted to discussing the scenario of the extremal black hole corresponding to the Nariai solution, where the Universe horizon and the event horizon coincides.
In Section V, the quasinormal frequencies are evaluated by employing both the WKB approximation and the matrix method.
The effects of various parameters, such as the equation of state, spin-orbit quantum number, and electric charge on the resultant quasinormal modes, are investigated numerically.
In particular, the quasinormal frequencies are analyzed for the case of a second type of extremal black hole associated with the Nariai solution.
The temporal evolution for small initial disturbance is also calculated by the finite difference method.
In particular, we explore the late-time tails of quasinormal modes by choosing specific metrics reflecting different asymptotically spacetime properties.
In the last section, concluding remarks are presented in addition to further discussions.

\section{II. Spherically symmetric power-Maxwell charged black holes in Rastall Gravity}\label{section2}
According to Rastall gravity, Eq.~(\ref{N1}) inspires the following gravitational field equation
\begin{equation}\label{N2}
{R_{\mu \nu }} + \left(\kappa _0\lambda-\frac12\right) {g_{\mu \nu }}R = \kappa _0 {T_{\mu \nu }}.
\end{equation}
where ${\kappa _0}$ is a constant which, by taking contraction of the field equation, gives rise to an additional relation between the spacetime curvature and the trace anomaly of the energy-momentum tensor~\cite{prd-rastall-01}.
It can be readily verified that the theory falls back to Einstein's gravity by taking $\kappa_0 = 1$ and $\lambda=0$.
In this paper, we employ the spherically symmetric black hole metric recently proposed in Rastall theory~\cite{grg-rastall-hig-char-bh-11}.
The obtained metric is a generalization of various previous studies, of which we give a brief account as follows.
The static black hole solution in question is surrounded by quintessence fluid in the presence of a power-Maxwell field.
The present study will focus on the solution in four-dimensional spacetime, which reads
\begin{equation}\label{N3}
d{s^2} =  - f(r)d{t^2} + f{(r)^{ - 1}}d{r^2} + {r^2}d{\theta ^2} + {r^2}{\sin ^2}\theta d{\phi ^2}.
\end{equation}
By assuming that the electromagnetic field is spherically symmetric and static, its action can be written as~\cite{epj-charg-btz-high-01}
\begin{equation}\label{N4}
{{\cal L}_F} =  - {( - \xi {\cal F})^s},
\end{equation}
where ${\cal F} = {F_{\mu \nu }}{F^{\mu v}}{\rm{ }}$, and ${F_{\mu \nu }}$ is antisymmetric Faraday tensor.
Here, $\xi $ and the exponent $s$ are constants.
To proceed, one evaluates the energy-momentum tensor of the electromagnetic field and the quintessence fluid
\begin{equation}\label{N5}
T_v^\mu  = E_v^\mu  + T_v^{*\mu },
\end{equation}
where, the Maxwell's stress tensor $E_v^\mu$ is found to be
\begin{equation}\label{N6}
E_\mu ^v =  - {( - \xi )^s}{({\cal F})^{s - 1}}\left( {2s{F_{\sigma \mu }}{F^{\sigma v}} - \frac{1}{2}\delta _\mu ^v{\cal F}} \right).
\end{equation}
$T_v^{*\mu }$ represents the energy-momentum tensor of the quintessence fluid.
By assuming the barotropic equation of state, namely, $p = w\rho $, one obtains~\cite{epjc-rastall-sclar-hig-quinte-13}
\begin{equation}\label{N7}
T_t^{*t} = T_r^{*r} =  - \rho (r),T_\theta ^{*\theta } = T_\varphi ^{*\varphi } = \frac{1}{{n - 1}}\rho (r)(3w + 1).
\end{equation}
By substituting the specific forms of the energy-momentum tensors, Eqs.~(\ref{N6}-\ref{N7}) into the Rastall gravitational field equation Eq.~({\ref{N2}}).
The resultant black hole solution reads~\cite{grg-rastall-hig-char-bh-11}
\begin{equation}\label{N10}
\begin{array}{l}
f(r) = 1 - \frac{{8\Gamma \left( {\frac{3}{2}} \right)M}}{{2{\pi ^{\frac{3}{2} - 1}}{r^{n - 2}}}} + \frac{{4{Q^2}\Gamma \left( {\frac{3}{2}} \right){{(2s - 1)}^{\frac{{3 - 2s}}{{1 - 2s}}}}{r^{\frac{2}{{1 - 2s}}}}}}{{{\pi ^{\frac{n}{2} - 1}}(n - 1)s(n - 2s)}}\\
 - \frac{{2\kappa _0 {C_a}{{(3(2\kappa _0 \lambda (w + 1) - 1) + 1)}^2}}}{{2n(8\kappa _0 \lambda  - 2)}} \times \frac{{{r^{ - {r_B}}}}}{{2w - 2\kappa _0 \lambda (w + 1)}} + \frac{{2\Lambda {r^2}}}{{3(8\kappa _0 \lambda  - 2)}}.
\end{array}
\end{equation}
where ${r_B} = \frac{{2(n(w + 1)(2\kappa _0 \lambda  - 1) + 2)}}{{3(2\kappa _0 \lambda (w + 1) - 1) + 1}}$, ${C_a}$ is a constant of integration.
$M$ is related to the mass of the black hole and $\Lambda $ is the cosmological constant.
Among them, the value of $\omega$, associated with the properties of the equation of state of the quintessence field, is constrained by cosmological observations and is found to $\sim - 1$~\cite{book-weinberg-cosmology}.
Also, $n$ is the number of spatial dimensions of the spacetime in question, and we will only consider the case $n=3$ in our present study.

As mentioned above, even for asymptotically flat spacetimes, the black hole metrics in Rastall gravity might behave asymptotically as those of de Sitter spacetimes.
Thus the motivation of the present study, in part, is to investigate whether the late-time tail in such a metric still bears the same features established in Einstein gravity.
For this purpose, the following choices of parameters are mostly adopted for the remainder of the work.
First, one sets $\Lambda=0$ so that the original spacetimes are indeed asymptotically flat.
Moreover, we assume $\kappa_0=1$ and $C_a=-1$ in order to establish a black hole metric solution that meets the above requirement.
To be more specific, this choice is simple enough; meanwhile, it still guarantees a nonvanishing contribution from the fourth term on the r.h.s. of Eq.~\eqref{N10}.
As discussed below, it eventually leads to the metrics whose spacetimes are asymptotically de Sitter and thus suffices for our purpose.
Besides, we choose $\lambda = 1$ in order to simplify the forms of resultant metrics further.
Lastly, in particular cases, the above parameter can be varied to investigate the properties of the resultant quasinormal frequencies as the theory further deviates from the corresponding Einstein gravity by taking $\lambda=0$.

For linear Maxwell field electromagnetic field, $s = 1$.
The metric is simplified to read
\begin{align}\label{N11}
{f_{1*}}(r) =  1 + \frac{{{Q^2}}}{{{r^2}}} - \frac{{2M}}{r}
 -r^{-\frac{3w+5}{3w+2}} \frac{\left(3w+2\right)^2}{9}.
\end{align}
In the case of the nonlinear electromagnetic field, we will explicitly consider the cases $s = 2$ and $s = 3$.
Accordingly, the metric takes the forms
\begin{align}\label{N12}
{f_{2*}}(r) =  1 - \frac{{2M}}{r} - \frac{{{3^{1/3}}{Q^2}}}{{2{r^{2/3}}}}
 -r^{-\frac{3w+5}{3w+2}} \frac{\left(3w+2\right)^2}{9},
\end{align}
and
\begin{align}\label{N13}
{f_{3*}}(r) = 1 - \frac{{2M}}{r} - \frac{{{5^{3/5}}{Q^2}}}{{9{r^{2/5}}}}
 -r^{-\frac{3w+5}{3w+2}} \frac{\left(3w+2\right)^2}{9}.
\end{align}

\section{III. The master equation for massless Dirac spinor}\label{section3}

The Dirac equation for a massless spinor in curved spacetime can be written as~\cite{rmp-neutri-gravi-inter-01}
\begin{equation}\label{N14}
[{\gamma ^a}e_a^\mu \left( {{\partial _\mu } + {\Gamma _\mu }} \right)]\Psi  = 0,
\end{equation}
where ${\gamma ^a}$ are the gamma matrices.
\begin{equation}\label{N15}
{\gamma ^0} = \left( {\begin{array}{*{20}{c}}
{ - i}&0\\
0&i
\end{array}} \right),\quad {\gamma ^i} = \left( {\begin{array}{*{20}{c}}
0&{ - i{\sigma ^i}}\\
{i{\sigma ^i}}&0
\end{array}} \right),\quad i = 1,2,3.
\end{equation}
where ${\sigma ^i}$ are the Pauli matrices.
${\Gamma _\mu }$ is spin connection coefficient which is defined as
\begin{equation}\label{N16}
{\Gamma _\mu } = \frac{1}{8}\left[ {{\gamma ^a},{\gamma ^b}} \right]e_a^v{e_{bv;\mu }}.
\end{equation}
$e_\mu ^a$ is known as tetrad, a set of orthogonal space time vector fields that satisfies
\begin{equation}\label{N17}
{g_{\mu v}} = {\eta _{ab}}e_\mu ^ae_v^b,
\end{equation}
where ${\eta _{ab}} = {\mathop{\rm diag}\nolimits} ( - 1,1,1,1)$ is the Minkowski metric.
Now, the static spherically symmetric black hole metric, Eq.~(\ref{N3}), implies
\begin{equation}\label{N19}
e_\mu ^a = {\mathop{\rm diag}\nolimits} \left( {f{{(r)}^{1/2}},f{{(r)}^{ - 1/2}},r,r\sin \theta } \right).
\end{equation}
By rescaling the wave function
\begin{equation}\label{N20}
\Psi  = f{(r)^{ - \frac{1}{4}}}\Phi.
\end{equation}
The Dirac equation can be rewritten in terms of $\Phi$, which reads
\begin{equation}\label{N21}
\begin{array}{l}
\left[ {{\gamma ^0}f{{(r)}^{ - 1/2}}\frac{\partial }{{\partial t}} + {\gamma ^1}f{{(r)}^{1/2}}\left( {\frac{\partial }{{\partial r}} + \frac{1}{r}} \right)} \right.
\left. { + {\gamma ^2}\frac{1}{r}\left( {\frac{\partial }{{\partial \theta }} + \frac{1}{2}\cot \theta } \right) + {\gamma ^3}\frac{1}{{r\sin \theta }}\frac{\partial }{{\partial \phi }}} \right]\Phi  = 0.
\end{array}
\end{equation}
The above equation can be solved by using the method of separation of variables.
By factoring out the angular part of the wave function according to standard procedure~\cite{rmp-neutri-gravi-inter-01,book-relati-ele-theo,jpa-unifi-nonrelati-relati-gree,jpa-unifi-nonrelati-relati-redu-gree,jpa-unifi-nonrelati-relati-wavefuc}, one obtains the resultant master equations regarding the radial part.

Furthermore, the upper and lower components of the spinnor, $F^{(\pm )}$ and $G^{(\pm )}$, can be decoupled and rewritten as
\begin{equation}\label{N26}
\begin{array}{l}
\left( { - \frac{{{d^2}}}{{dr_*^2}} + {V_{(\pm )1}}} \right){F^{(\pm )}} = {\omega ^2}{F^{(\pm )}},\\
\left( { - \frac{{{d^2}}}{{dr_*^2}} + {V_{(\pm )2}}} \right){G^{(\pm)}} = {\omega ^2}{G^{( \pm)}},
\end{array}
\end{equation}
where
\begin{equation}\label{N27}
{V_{( \pm )1}} =   \frac{{d{W_{( \pm )}}}}{{d{r_*}}} + W_{( \pm)}^2 , ~{V_{( \pm )2}} =  - \frac{{d{W_{( \pm )}}}}{{d{r_*}}} + W_{( \pm)}^2,
\end{equation}
and
\begin{equation}\label{N28}
{W_ \pm } = \sqrt {f(r)} \frac{{\left| {{\kappa_ \pm }} \right|}}{r} ,
\end{equation}
where
\begin{equation}
{\kappa_\pm} =
\left\{
\begin{array}{*{20}{c}}
- (j + 1/2),&{j = l + 1/2}\\
{(j + 1/2),}&{j = l - 1/2}.
\end{array}
\right.
\end{equation}

It can be inferred from Eq.~({\ref{N27}}) that the effective potential ${V_{(\pm)1,2}}$ is related to the equation of state via $w$ and the charge $Q$ of the background metric, as well as to the spin-orbit quantum number $\kappa_\pm$ of the wave function.
As discussed in Refs.~\cite{prd-Intertwi-equa-bh-pertu,prd-dirac-qnm-sch}, the master equations for the upper and lower components lead to the identical spectra of quasinormal frequencies, since the potentials ${V_{(\pm)1,2}}$ are supersymmetric partners.
In particular, the two forms of the potentials defined in Eq.~\eqref{N27} are related to each other through the Darboux transformation.
As pointed out in Ref.~\cite{Konoplya:2019xmn,Konoplya:2019ppy}, by the WKB method, the master equation for the lower component leads to better accuracy.
Therefore, in what follows, we will concentrate on the master equation for $G$ and investigate the associated quasinormal frequencies, as well as their dependence on various physical parameters.

\section{IV. Two types of extremal black hole solutions and the associated charges}\label{section4}
In this section, we demonstrate that for specific scenarios, the metric possesses two types of extremal black hole solutions.
Besides conventional one found in the Reissner-Nordstr\"{o}m metric, the second type of extremal black hole solution is associated with the Nariai solution~\cite{grg-nariai,prd-quantum-evolution-nariai,ijmpa-nariai-sch-desitter}.
The related quasinormal frequencies will be studied in the next Section.

\begin{figure*}[htbp]
\centering
\includegraphics[scale=0.33]{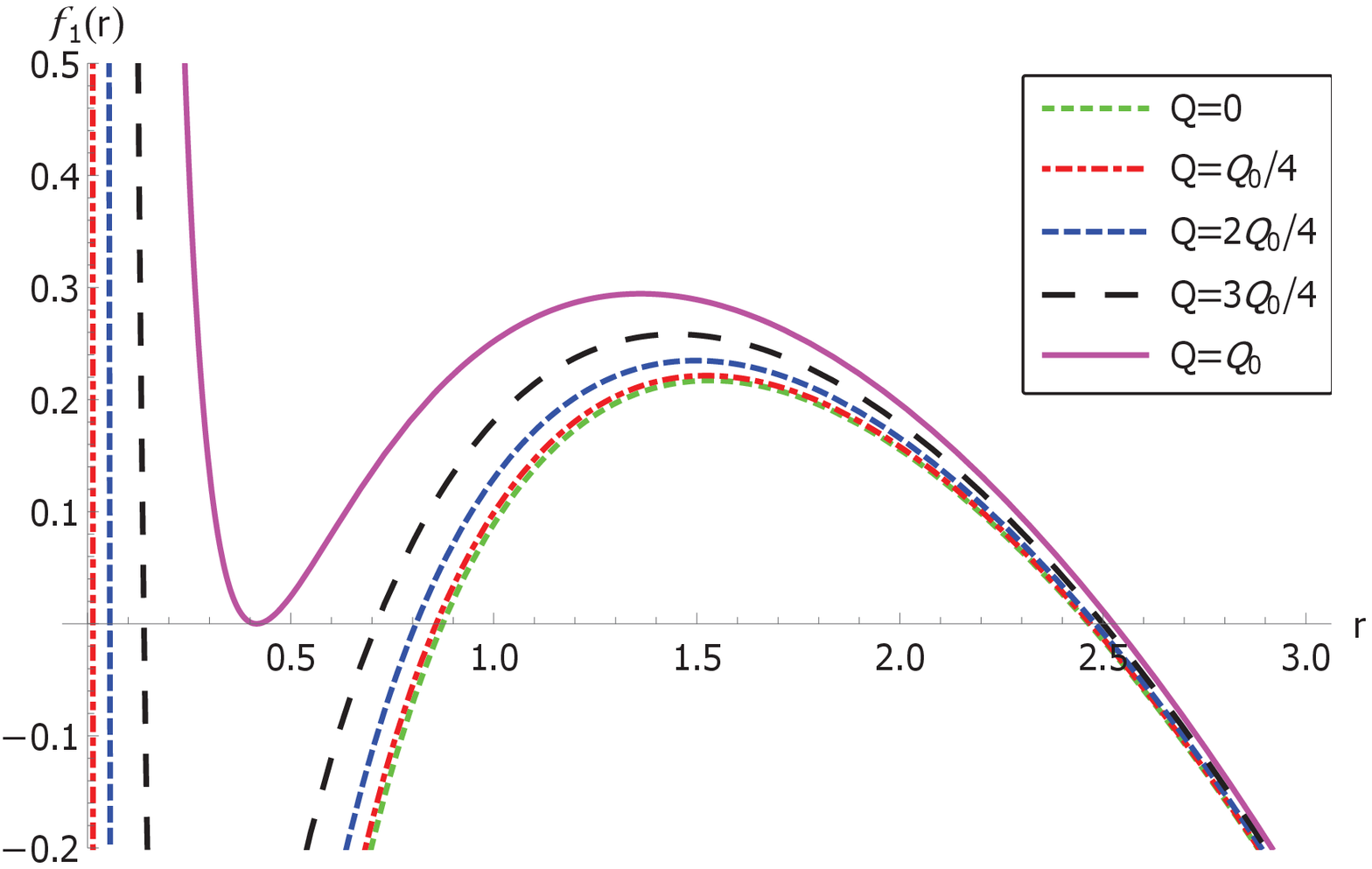}
\includegraphics[scale=0.33]{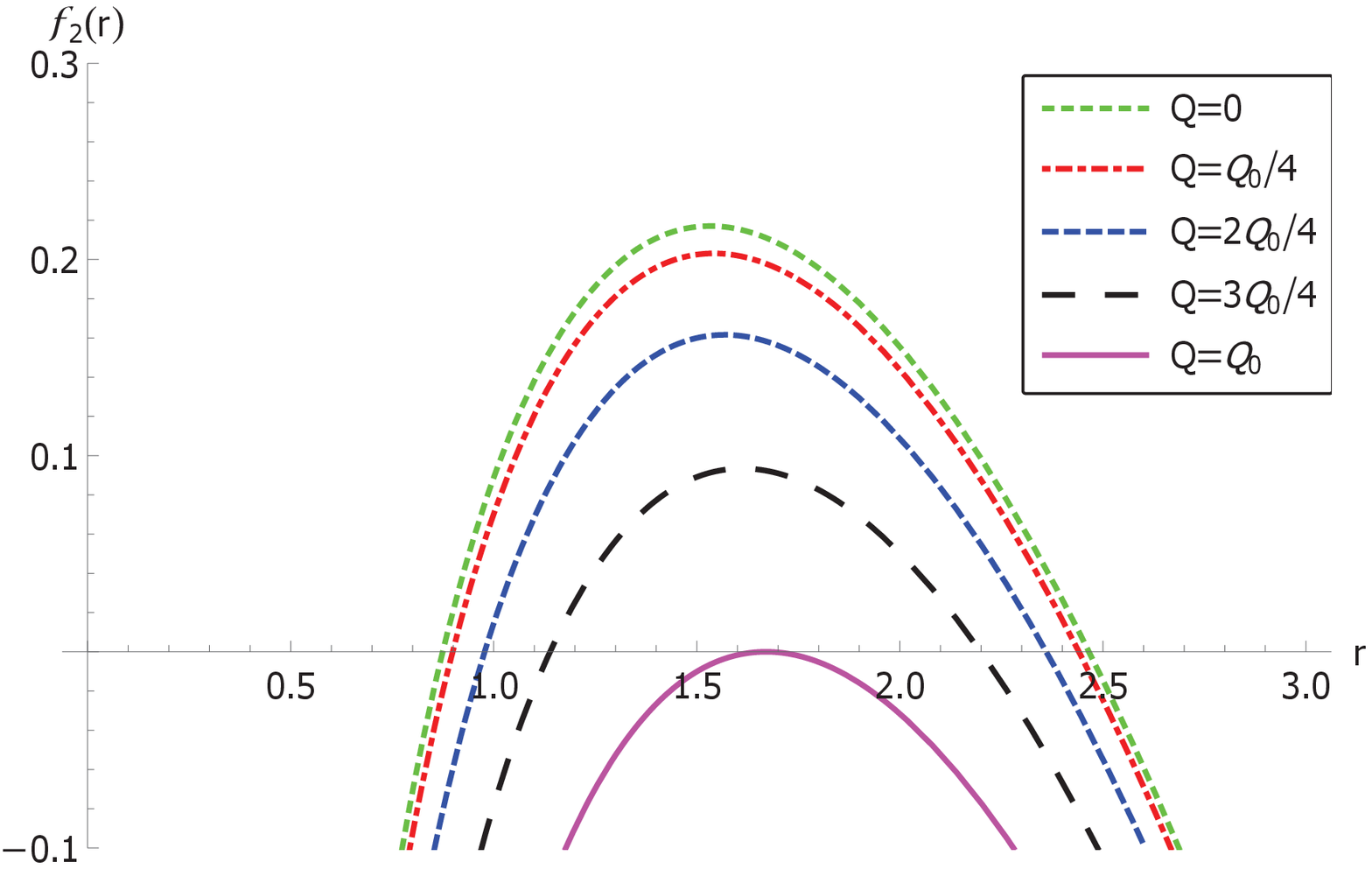}
\includegraphics[scale=0.33]{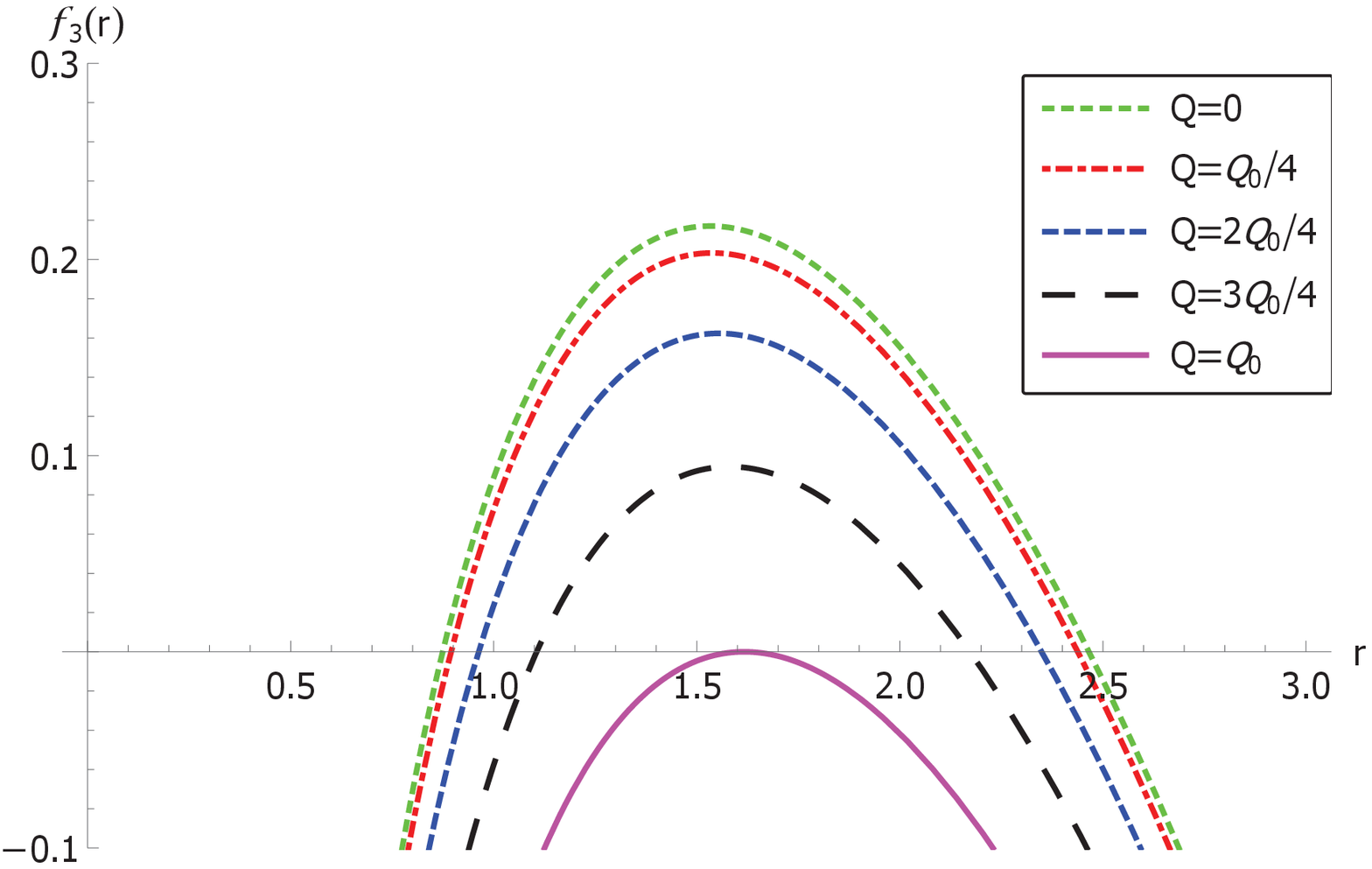}
\caption{(Color Online) The functions ${f_1}(r)$ (left), ${f_2}(r)$ (middle), and ${f_3}(r)$ (right) vs. $r$ evaluated for given $w=-1$}\label{Fig1}
\end{figure*}

We first study the case of linear electromagnetic field with $s = 1$ and quintessence fluid with $w  =  - 1$.
The resulting metric can be readily obtained by Eq.~(\ref{N11}), which reads
\begin{equation}\label{N32}
{f_1}(r) = 1 + \frac{{{Q^2}}}{{{r^2}}} - \frac{{2M}}{r} - \frac{{{r^2}}}{9} .
\end{equation}
By excluding one negative root of Eq.~(\ref{N32}) and enumerating the rest, one finds that the metric implies three horizons.
They are the inner horizon $r_i$, event horizon $r_h$, and cosmological horizon $r_c$, which satisfy $r_i \le r_h \le r_c$ as shown in Fig.~\ref{Fig1}.
In this case, one observes the extremal solution in the same context of the Reissner-Nordstr\"{o}m black hole.
Here, one may consider an imaginary process as the charge of the black hole increases while the mass remains unchanged, the extremal black hole takes place when the inner horizon coincides with the event horizon.
If one further adds more charge to the black hole, the horizon in question disappears and, therefore, as a physical process, prohibited by the cosmic censorship hypothesis.
An explicit example is given in Fig.~\ref{Fig1}.
There, one finds the resultant extremal value ${Q_0} = 0.4038212447185102$ and ${r_i} = {r_h} = 0.41599780536901554$ for an arbitrarily chosen mass $M=0.4$.
In particular, it is convenient to rewrite the charge and mass $Q$ and $M$ in terms of $r_i$ and $r_h$.
It is not difficult to find that the resultant relations are
\begin{equation}\label{N33}
\begin{array}{l}
M = \frac{1}{{18}}\left( {9{r_h} - {r_h}^3 + 9{r_i} - {r_h}^2{r_i} - {r_h}{r_i}^2 - {r_i}^3} \right)
\end{array}
\end{equation}
and
\begin{equation}\label{N34}
\begin{array}{l}
Q = \frac{1}{3}\sqrt {{r_h}} \sqrt {{r_i}} \sqrt {9 - {r_h}^2 - {r_h}{r_i} - {r_i}^2}.
\end{array}
\end{equation}
One can evaluate the condition for extremal black hole by substituting $r_i=r_h$ into Eqs.~(\ref{N33}-\ref{N34}) and solves for $Q=Q_0$ and $r_i$ with given $M$.
This result is presented by the solid magenta curve in the left plot of Fig.~\ref{Fig1}.

Next, let us consider the cases for the nonlinear electromagnetic field with $s = 2$ and $s = 3$.
Similarly, the resultant metrics are found to be
\begin{equation}\label{N35}
{f_2}(r) = 1 - \frac{{2M}}{r} - \frac{{{3^{1/3}}{Q^2}}}{{2{r^{2/3}}}} - \frac{{{r^2}}}{9}
\end{equation}
and
\begin{equation}\label{N38}
{f_3}(r) = 1 - \frac{{2M}}{r} - \frac{{{5^{3/5}}{Q^2}}}{{9{r^{2/5}}}} - \frac{{{r^2}}}{9} .
\end{equation}
As shown in middle and right plot of Fig.~\ref{Fig1}, when the charge $Q$ increases, a different behavior is observed.
Now the event horizon ${r_h}$ and cosmological horizon ${r_c}$ approach each other and eventually collide.
Again, we use ${Q_0}$ to denote the charge when the two horizons coincide.
One may imagine an observer who sits between the event and the cosmological horizon.
For the present case, his presence seems to be apparently rendered unattainable as the black hole solution becomes an extremal one.
In fact, a closer look reveals that the proper distance between the two horizons still remains finite during the process~\cite{mpla-nariai-quintes}.
In this case, the spacetime on the causal patch between the horizons is approximately that of the Nariai solution.
Obviously, this extremal solution is intrinsically different from the one discussed before, which is related to the so-called Nariai solution.
To proceed with the numerical study, again, we choose $M=0.4$.
For the metric $f={f_2}$, one finds ${Q_0} = 0.6418853593399654$, ${r_c} = {r_h} = 1.6700060557527228$.
For the metric $f={f_3}$, ${Q_0} = 0.944331289742155$, ${r_c} = {r_h} = 1.6166133524816748$ is obtained.
If one continues to increase the charge until it exceeds ${Q_0}$, both horizons will disappear.

It is readily to verify that metrics given in Eqs.~\eqref{N32}, \eqref{N35}, and \eqref{N38} are asymptotically de Sitter.
They will subsequently be employed in the numerical studies of quasinormal modes in the following section.

\section{V. Quasinormal frequencies for massless Dirac field}\label{section5}

In this section, we numerically study the quasinormal frequencies of the massless Dirac field by employing the WKB approximation, matrix method, and finite difference method.
Based on Eq.(\ref{N26}), the relevant master equation reads
\begin{equation}\label{N45}
\left( { - \frac{{{d^2}}}{{dr_*^2}} + V(r,\kappa)} \right){G^{(\pm )}} = {\omega ^2}{G^{(\pm )}},
\end{equation}
where
\begin{equation}\label{N46}
\begin{array}{l}
V(r,\kappa) =  -\frac{{d{W_ \pm }}}{{d{r_*}}} + W_ \pm ^2\\
{W_ \pm } = \sqrt {f(r)} \frac{{\left| {{\kappa_ \pm }} \right|}}{r}.
\end{array}
\end{equation}

First of all, by examing Eq.~\eqref{N3} and \eqref{N32}, it is straightforward to see that, for this particular form of the black hole solution is indeed equivalent to that in Reissner-Nordstrom-de Sitter spacetime for Einstein gravity.
To validate our numerical approach, we particularly adjust the metric parameters to match those discussed in Ref.~\cite{prd-RNds}.
Subsequently, numerical calculations are carried out and compared against the ones~\cite{prd-RNds} obtained by the Posh-Teller potential approximation.
The comparison indicates a reasonable agreement between the existing results and those by the WKB approximation and the matrix method, employed in the present study.

\begin{table*}[!t]
\caption{\label{tab1}
The calculated quasinormal frequencies $\omega $ for different metrics by using the third as well as sixth order WKB approximations and the matrix method.
The calculations are carried out with $ w=-49/50 $ and $\left| \kappa \right| = 2$.
In the first block, from top to bottom, the results are obtained by taking $\lambda=1$, $2$, and $3$.}
\newcommand{\tabincell}[2]{\begin{tabular}{@{}#1@{}}#2\end{tabular}}
\begin{ruledtabular}
\renewcommand\arraystretch{2}
\begin{tabular}{lllll}
$s$ &$Q$   & $\omega $(sixth order WKB)    & $\omega $(third order WKB)  & $\omega $(matrix method)       \\
\hline
 &${Q_0}/4$ & 0.731106-0.186040i & 0.730649-0.186079i & 0.733829-0.186571i \\
 &           &0.880836-0.224981i&0.880098-0.225072i&0.88550-0.229713i\\
  &           &0.916782-0.234420i&0.915966-0.234535i&0.921247-0.240282i\\
 1 &${Q_0}/2$  & 0.778436-0.192725i & 0.777910-0.192765i & 0.781392-0.194168i \\
 &           &0.919184-0.228683i&0.919973-0.228601i&0.923705-0.232527i\\
 &           &0.942553-0.240110i&0.952826-0.237409i&0.957167-0.241954i\\
  &$3{Q_0}/4$  & 0.874742-0.202474i & 0.874041-0.202505i & 0.878432-0.204115i \\
 &           &1.000408-0.232796i&1.001374-0.232739i&1.002931-0.236045i\\
 &           &0.931635-0.265624i&1.029349 -0.240239i&1.031297-0.243677i\\
  \hline
 &${Q_0}/4$ & 0.684335-0.175005i & 0.683936-0.175037i & 0.686996-0.174411i \\
2  &${Q_0}/2$  & 0.588898-0.149430i & 0.588623-0.149447i & 0.591668-0.147334i \\
  &$3{Q_0}/4$ & 0.427371-0.107361i & 0.427340-0.107353i & 0.429782-0.105460i \\
  \hline
 &${Q_0}/4$& 0.686898-0.186040i & 0.686494-0.186079i & 0.689531-0.174807i \\
 3 &${Q_0}/2$& 0.597226-0.150357i & 0.596964-0.150371i & 0.599823-0.148382i \\
  &$3{Q_0}/4$& 0.441336-0.108200i & 0.440039-0.108489i & 0.442261-0.106449i\\
\end{tabular}
\end{ruledtabular}
\end{table*}

\begin{figure*}[htbp]
\centering
\includegraphics[scale=0.45]{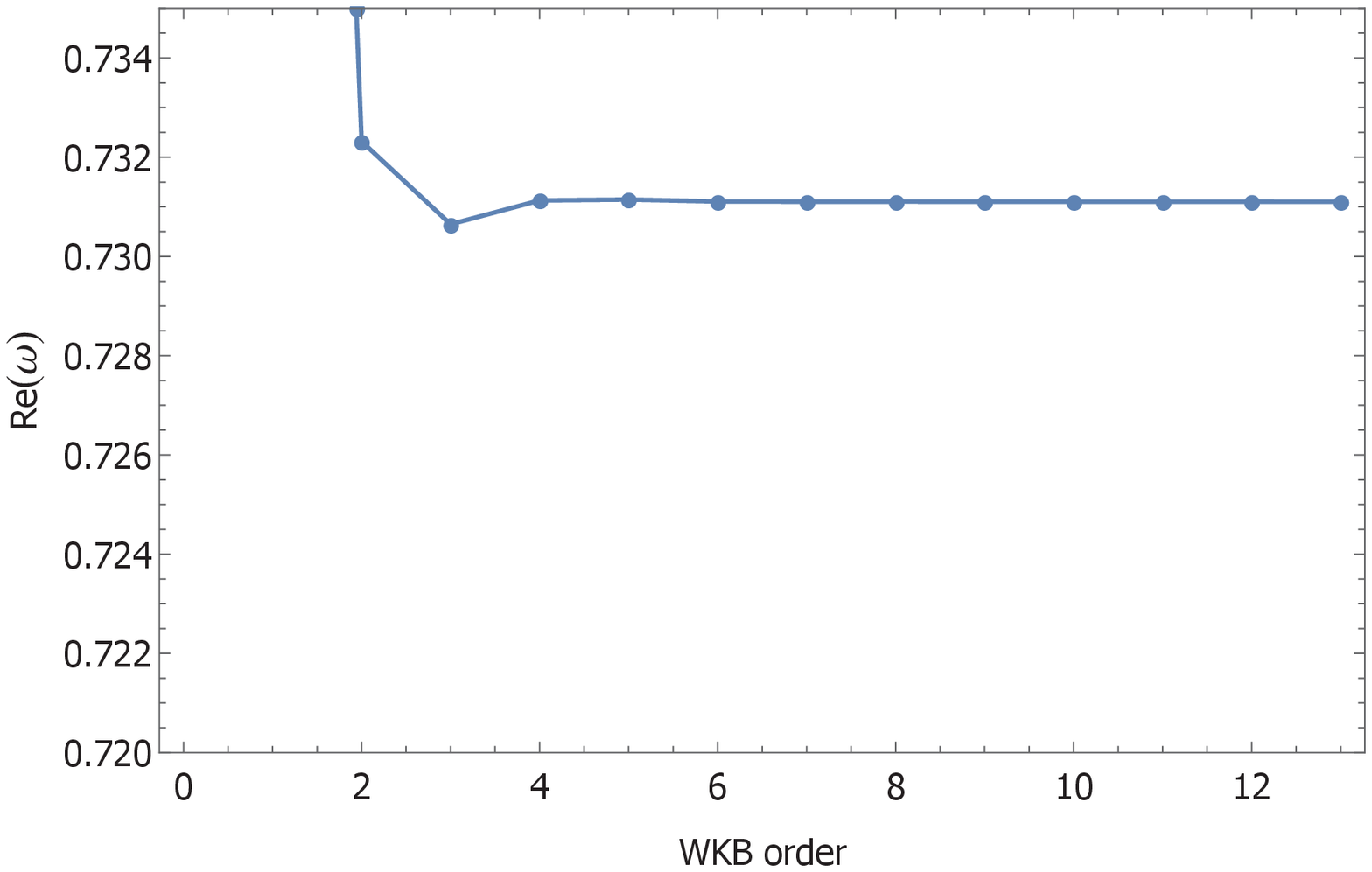}
\includegraphics[scale=0.45]{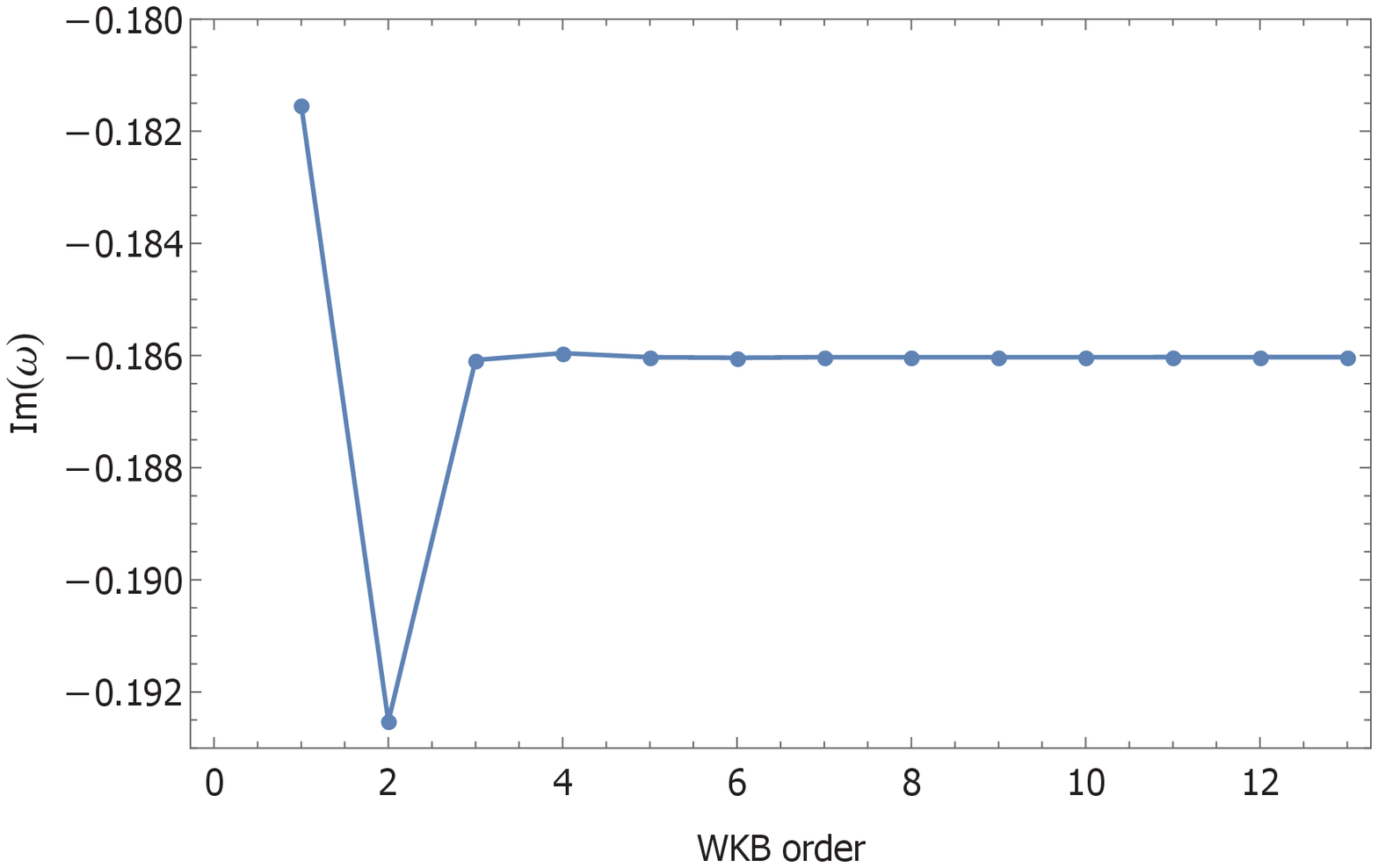}
\centering
\includegraphics[scale=0.45]{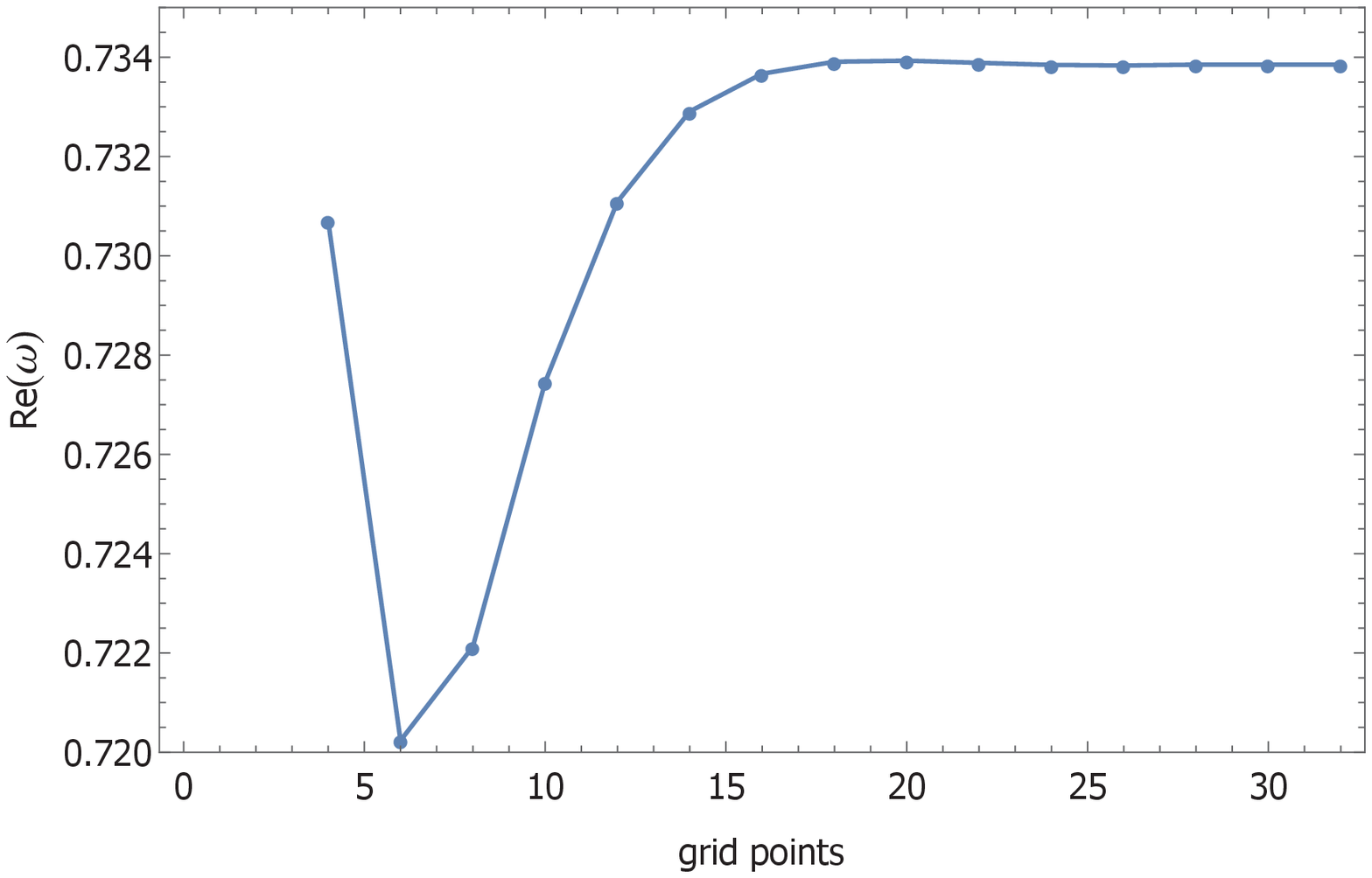}
\includegraphics[scale=0.45]{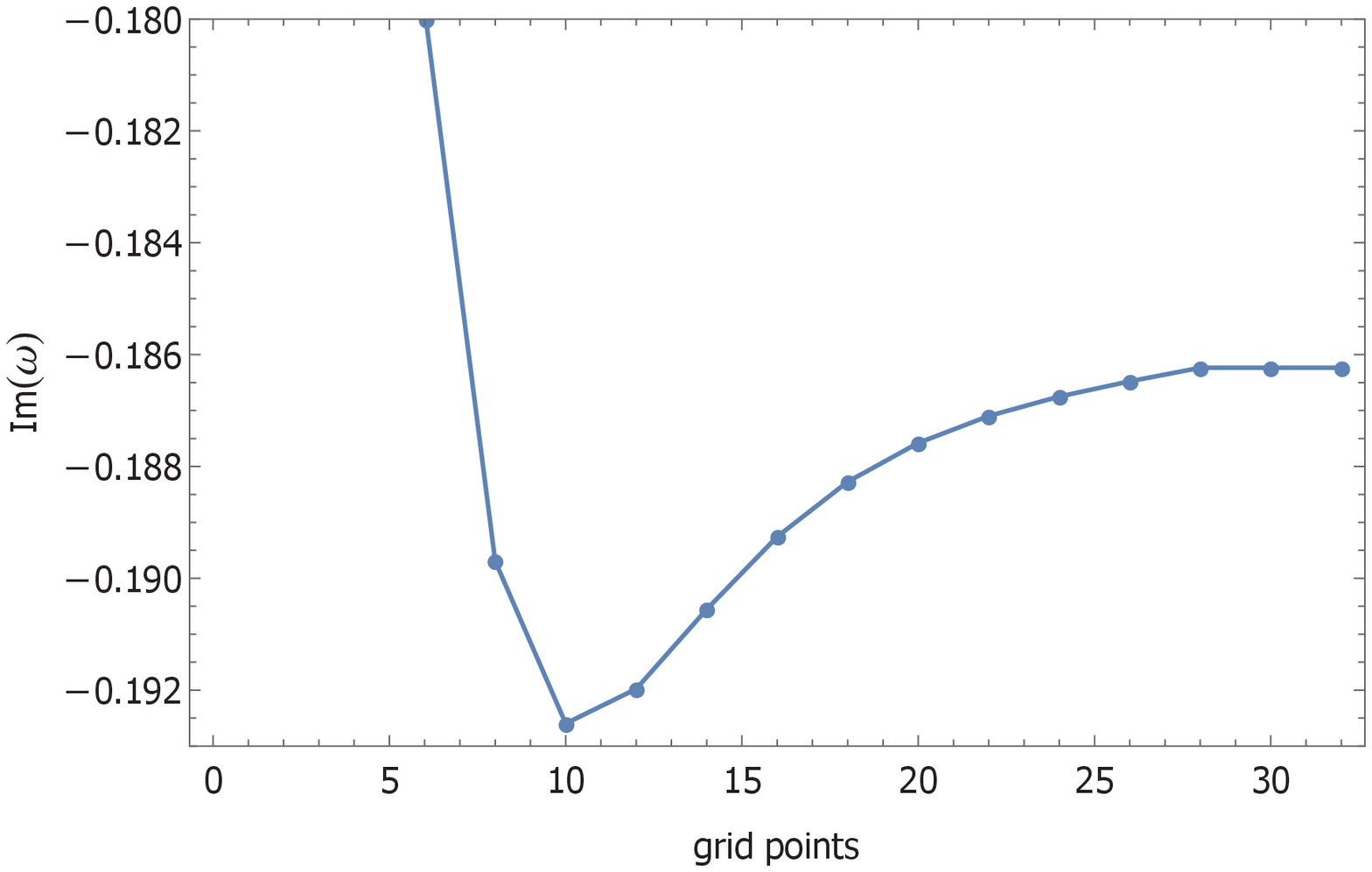}
\caption{(Color Online) The quasinormal frequencies of massless Dirac field obtained by the WKB method of different orders (top row) and those calculated by the matrix method with different numbers of grids (bottom row).
The calculations are carried out for the specific case with $w=-49/50$, $s=1$, $Q$=${Q_0}/4$, and $\left| \kappa \right| = 2$. }\label{Fig4}
\end{figure*}

Now, we proceed to present the resulting quasinormal frequencies obtained by using the third as well as sixth order WKB approximations in comparison to those evaluated by utilizing the matrix method.
The calculations are carried out by taking different values of $s$ and $Q$ and with given $ w=-49/50 $ and $\left| \kappa \right| = 2$.
The numerical results are presented in Tab.~\ref{tab1}.
It is found that the matrix method yields results consistent with those obtained by the WKB approximation.
The differences start to appear on the third significant figure, namely, less than one percent.
Moreover, a further study on the convergence of the methods employed in the present work is shown in Fig.~\ref{Fig4}.
There, the resultant quasinormal frequencies from the higher-order WKB method are based on original WKB formulae~\cite{Matyjasek:2017psv, Matyjasek:2019eeu, Konoplya:2019hlu}.
It is demonstrated that both methods are convergent as the calculated value gradually approaches a limit when the order or grid number increases.
For both cases, as a specific value is reached, the results can be considered as reliable.
In particular, for the WKB method, the obtained results are mostly stable for $n\ge 4$.
In the case of the matrix method, one should take a grid number more significant than $25$ for convergent results.

As discussed above, the parameter $\lambda$ measures the deviation of the theory from the Einstein gravity.
One may gradually vary this parameter in order to investigate the properties of the resultant quasinormal frequencies.
The results are shown in the first block of Tab.~\ref{tab1} where $s=1$.
In particular, from top to bottom, we take $\lambda=1, 2$, and $=3$ respectively while remaining other metric parameters unchanged.
The numerical results show that as the theory further deviates from the Einstein gravity, the real and the imaginary parts of the quasinormal modes both increase.
In other words, the damping for the Dirac field becomes faster as the Rastall parameter increases.
This result is different from the effect observed previously for massless scalar, electromagnetic, as well as gravitational fields~\cite{ctp-qnm-rastall}.

\begin{figure*}[htbp]
\centering
\includegraphics[scale=0.45]{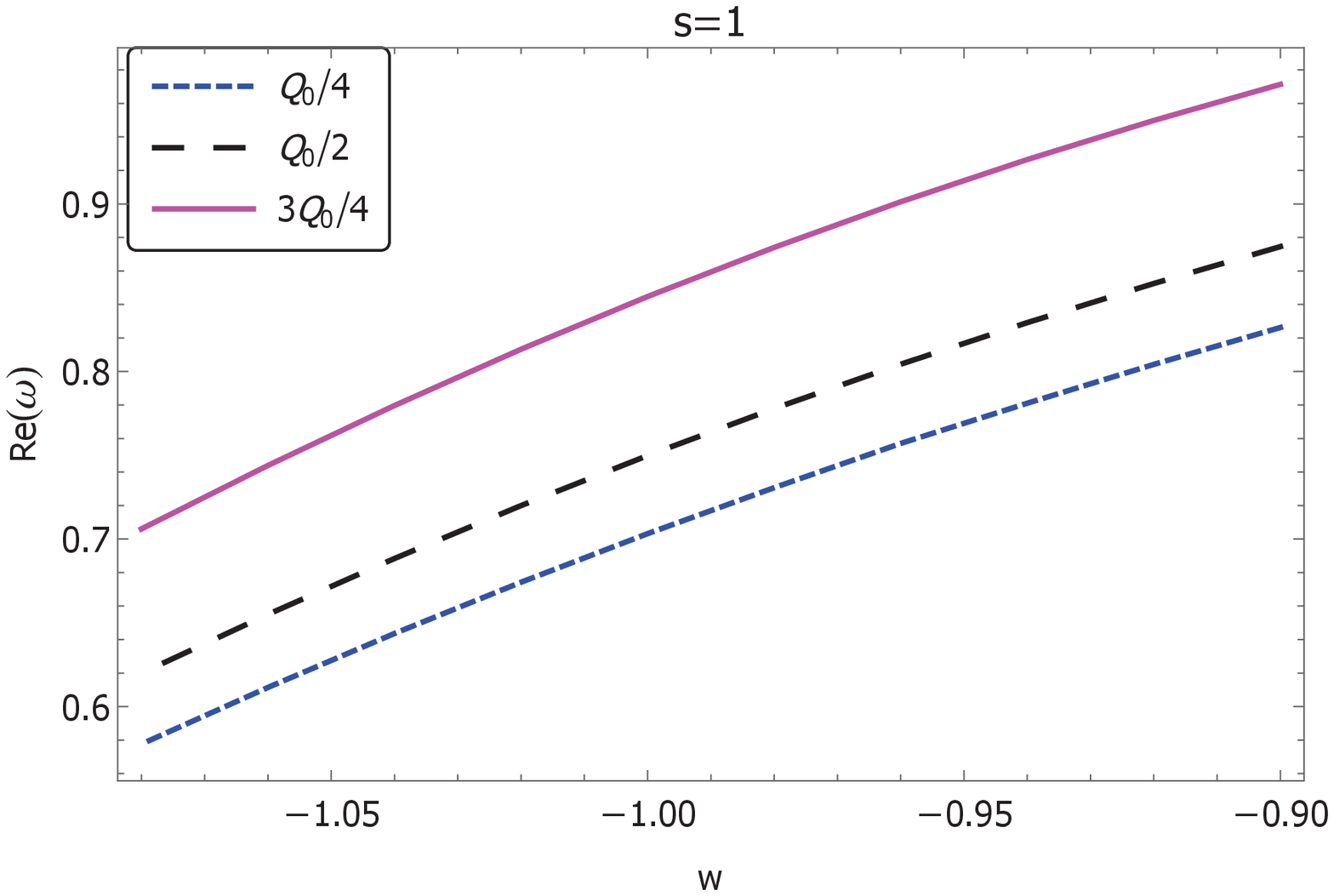}
\includegraphics[scale=0.45]{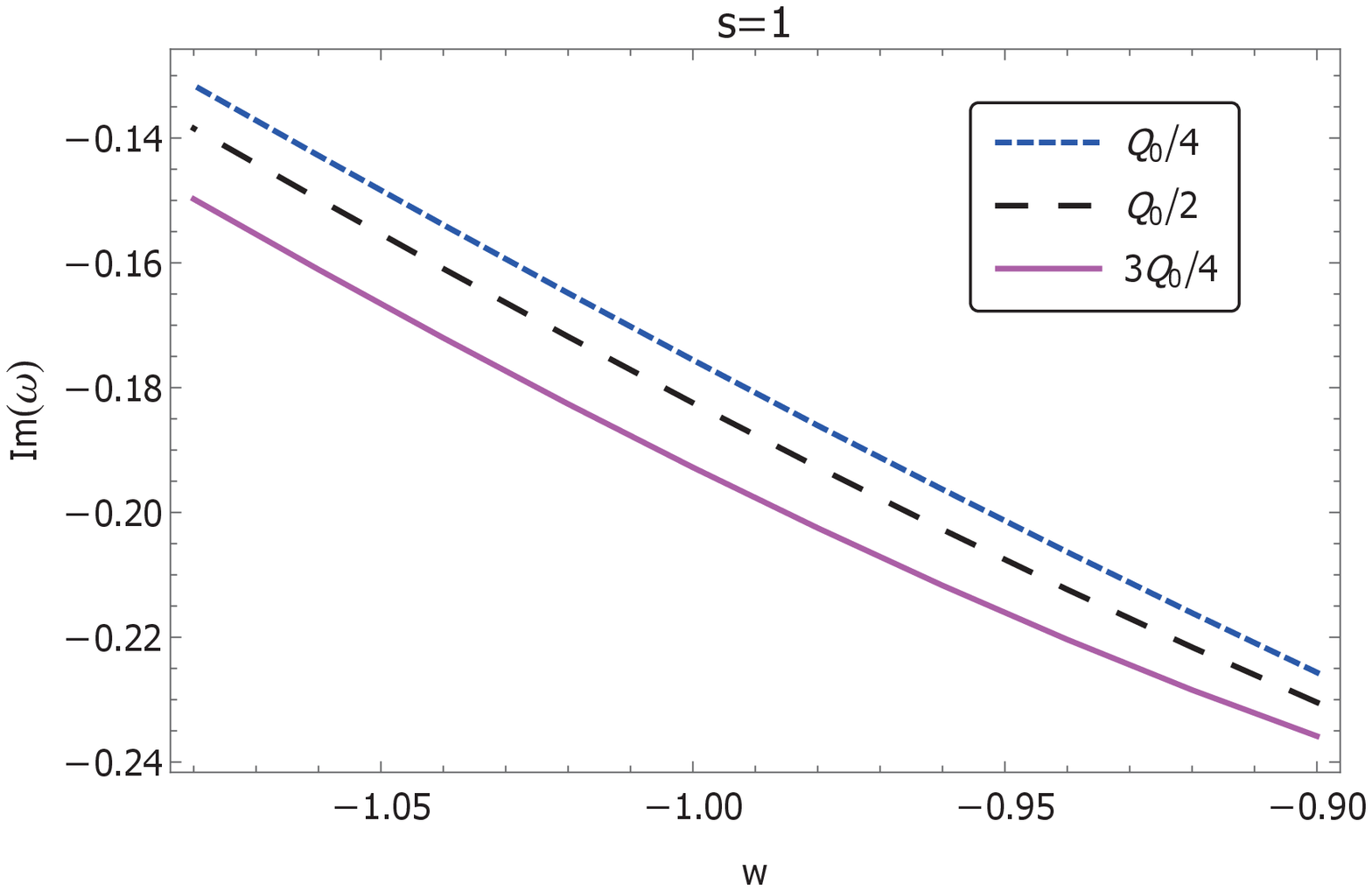}
\centering
\includegraphics[scale=0.45]{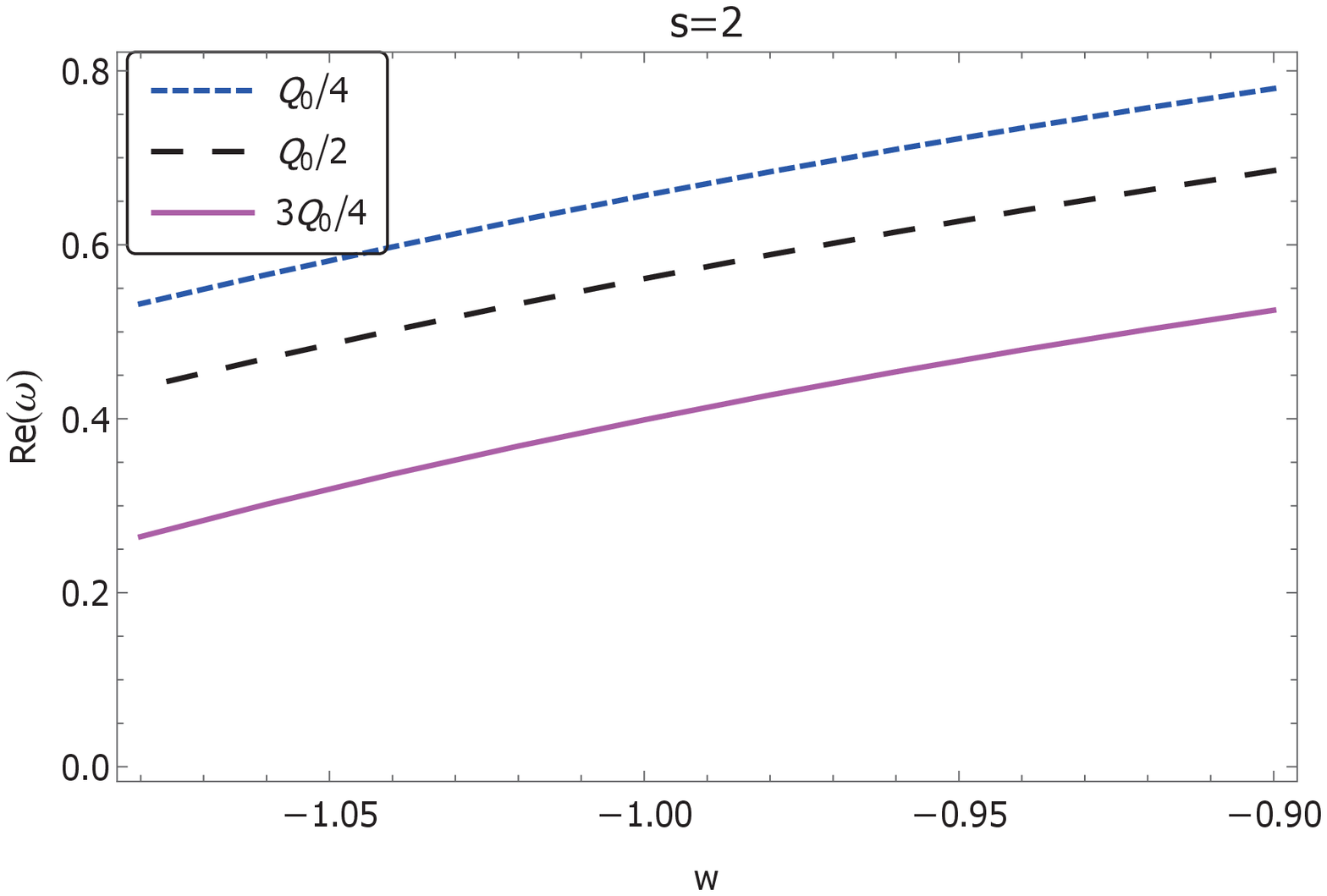}
\includegraphics[scale=0.45]{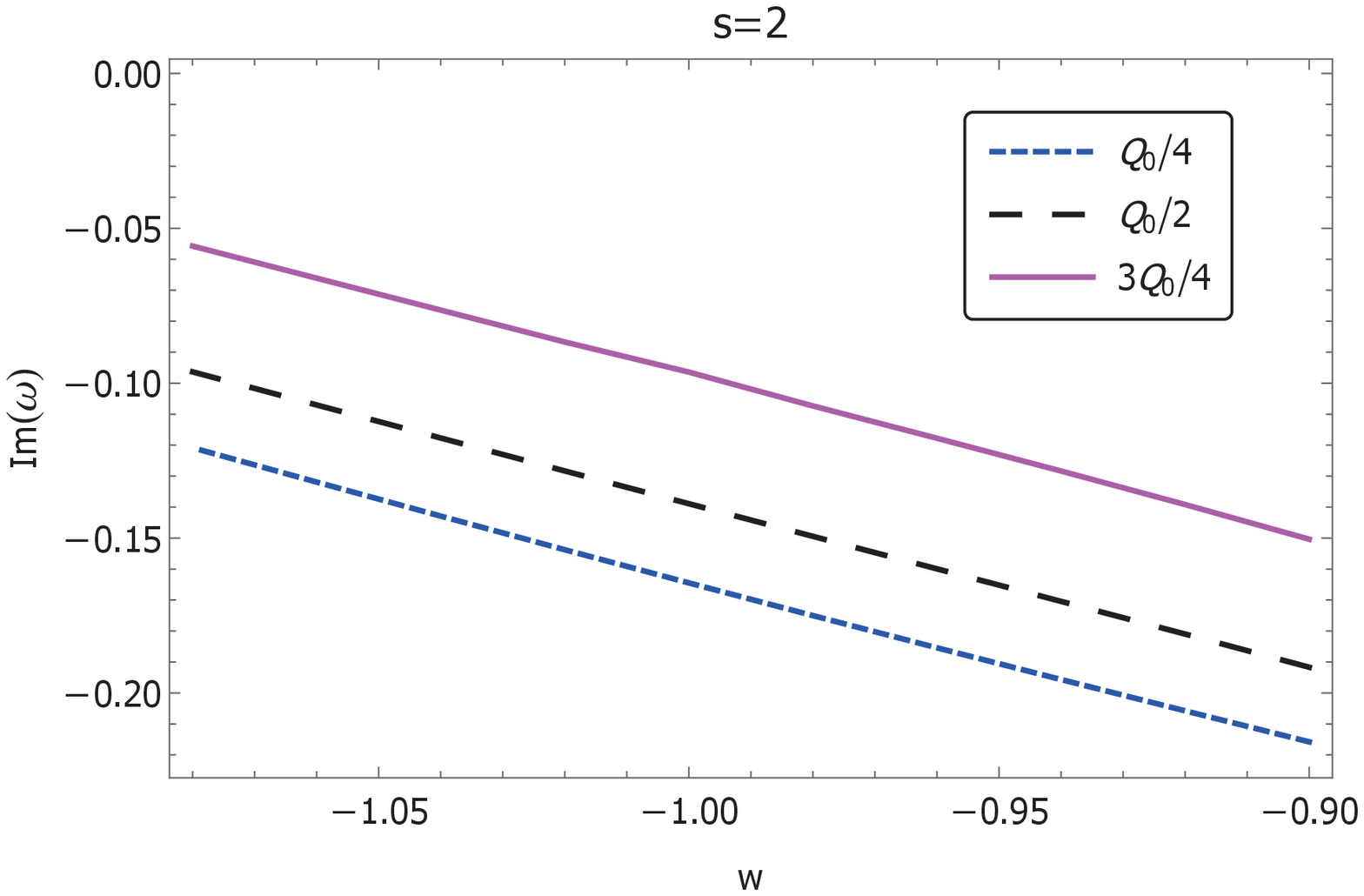}
\centering
\includegraphics[scale=0.45]{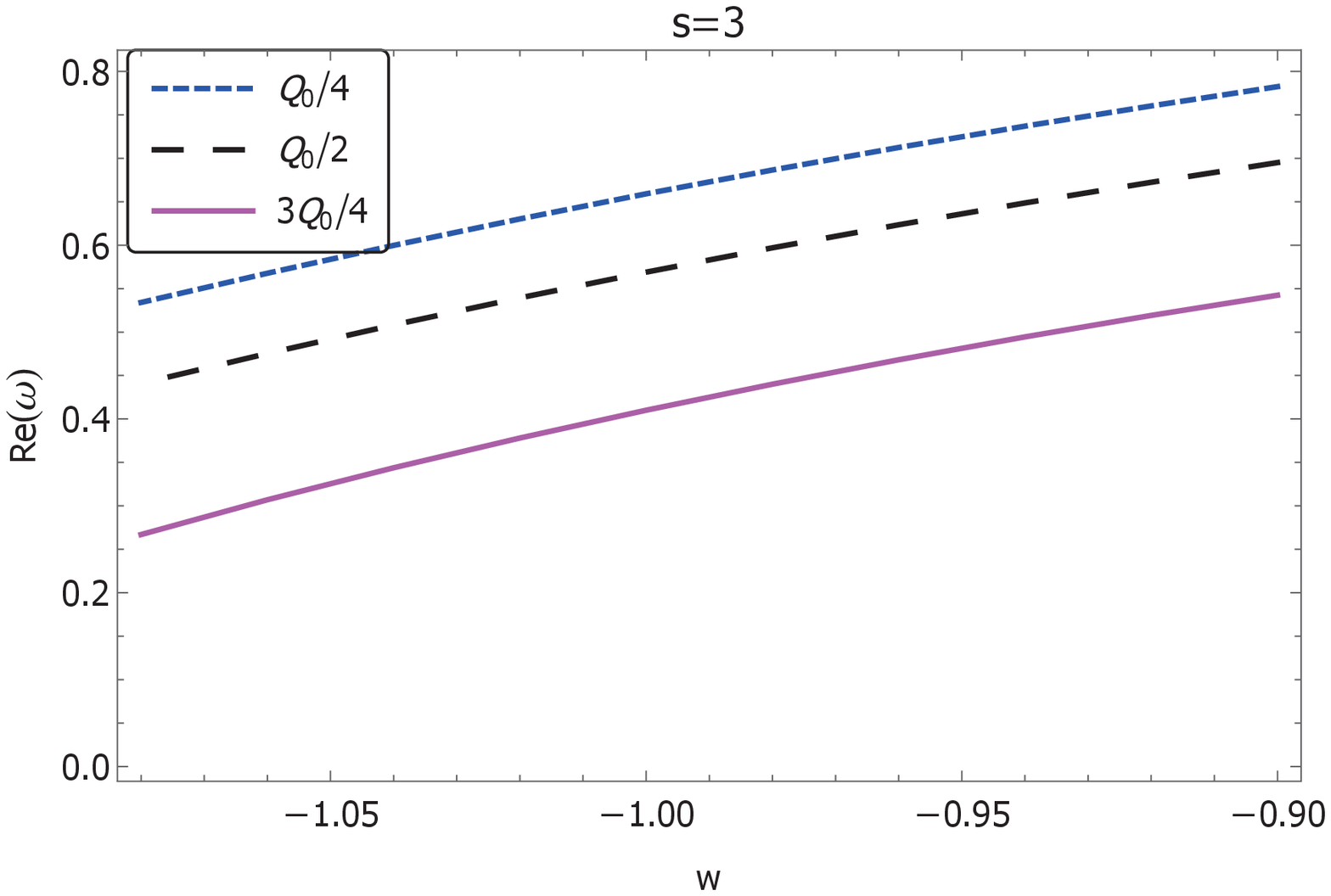}
\includegraphics[scale=0.45]{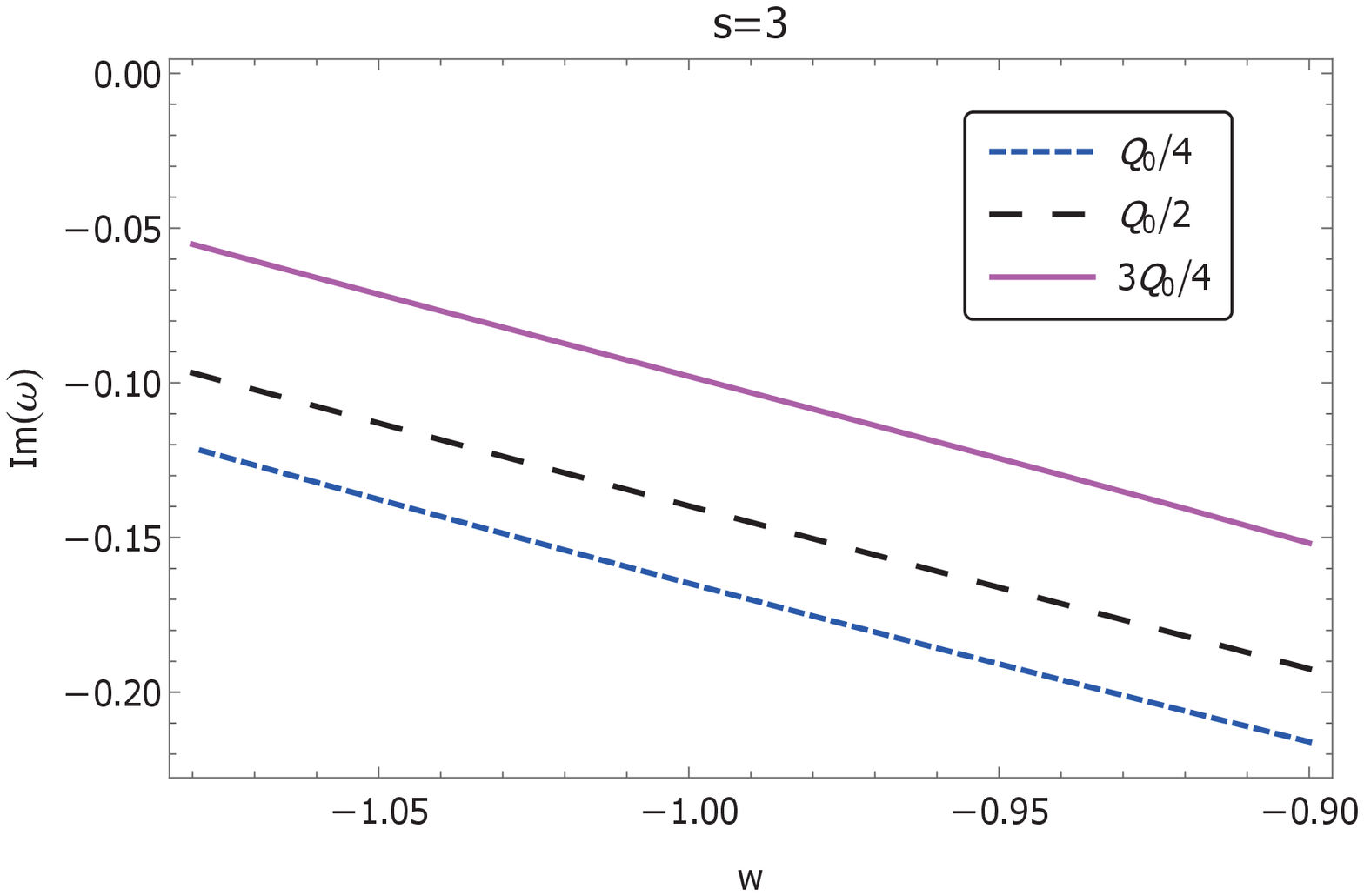}
\caption{(Color Online) The calculated quasinormal frequencies of massless Dirac field as functions of $w$.}\label{Fig3}
\end{figure*}

Next, we proceed to the study of the dependence of the evaluated quasinormal frequencies on the model parameters.
In Fig.~\ref{Fig3}, one depicts the quasinormal frequency of the Dirac field perturbation for different metrics parameters, $s$ and $Q$ as functions of $w$.
In all cases with given $s$ and $Q$, the real part of the quasinormal frequencies is found to monotonically increase with increasing $w$, while the imaginary part decreases with increasing $w$.
Moreover, the slope of the above curves is found to be mainly linear, independent of the value of the charge $Q$.
For $s=1$ and given $w$, the real part of calculated frequency increases as the charge $Q$ increases.
The imaginary part shows the opposite characteristic. It decreases with increasing charge.
For $s=2$ and $3$, on the other hand, the real part of calculated frequency decreases with increasing $Q$ for given $w$.
The imaginary part again shows the opposite characteristic, as it increases with increasing charge.
This indicates that the magnitudes of the quasinormal modes sensitively dependent on the equation of state of the surrounding matter, which is the quintessence fluid for the present case.
When the surrounding matter possesses a stiffer equation of state, the period of oscillations becomes more prolonged, while it decays faster.
This is somewhat intuitive as one might imagine the small initial perturbations have to push through more rigid matter while the oscillations lose the energy more rapidly.
The effect of charge on quasinormal modes is found to be opposite for black holes with linear and non-linear electromagnetic fields.

In Tab.~\ref{tab2}, we study the dependence of the quasinormal frequencies on the spin-orbit quantum number $\kappa$.
For the matrix method, we make use of 35 grid points.
It is understood that the accuracy will be improved further as the grid points increase.
From Tab.~\ref{tab2}, the real part of the quasinormal frequency increases as $|\kappa|$ increases.
While the imaginary part decreases with increasing $|\kappa|$.
This indicates that as $|\kappa|$ increases, the perturbation oscillates faster while it decays more slowly.

\begin{table*}[!t]
\caption{\label{tab2} The calculated quasinormal frequencies for different values of $|\kappa|$ obtained by the WKB approximation and the matrix method.
For the matrix method, 35 grid points are used.
Both calculations are carried out for the case of $n = 0$, $m = 0$, $w =  - 45/50$, and ${Q_{}} = {Q_0}/2$)}
\newcommand{\tabincell}[2]{\begin{tabular}{@{}#1@{}}#2\end{tabular}}
\begin{ruledtabular}
\renewcommand\arraystretch{2}
\begin{tabular}{llll}
$s$ & $|\kappa|$ & third order WKB     & matrix method                  \\
\hline
 & 1 & 0.433840915616-0.224966217228i & 0.432732727367-0.224998282790i \\
1  & 2 & 0.874530252970-0.230444389292i & 0.881776480990-0.225180338146i \\
  & 3 & 1.313643857600-0.231380833879i & 1.313196426297-0.230823673239i \\
   \hline
 & 1 & 0.345971291927-0.190840612924i & 0.352945552742-0.180621157207i \\
2  & 2 & 0.685255530112-0.191694452485i & 0.691175573306-0.190377632748i \\
  & 3 & 1.027427734248-0.192066710332i & 1.025420172252-0.190926933212i \\
   \hline
 & 1 & 0.350517288285-0.191322384727i & 0.351864569672-0.185715238599i \\
3  & 2 & 0.695018474234-0.192434284744i & 0.700624031577-0.190102313781i \\
  & 3 & 1.042181868101-0.192834699755i & 1.040130252800-0.191881255405i
\end{tabular}
\end{ruledtabular}
\end{table*}

\begin{figure*}[htbp]
\centering
\includegraphics[scale=0.45]{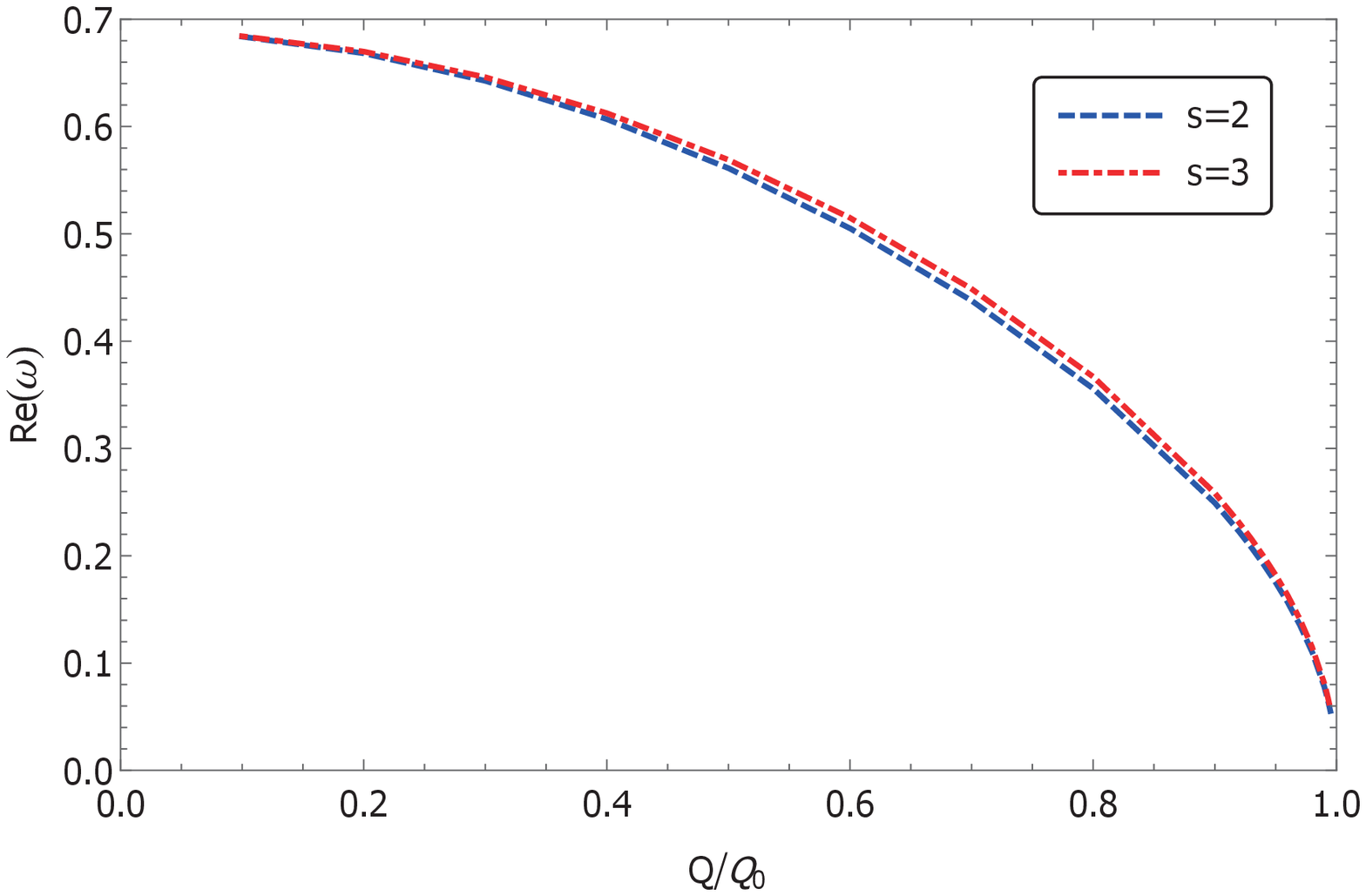}
\includegraphics[scale=0.45]{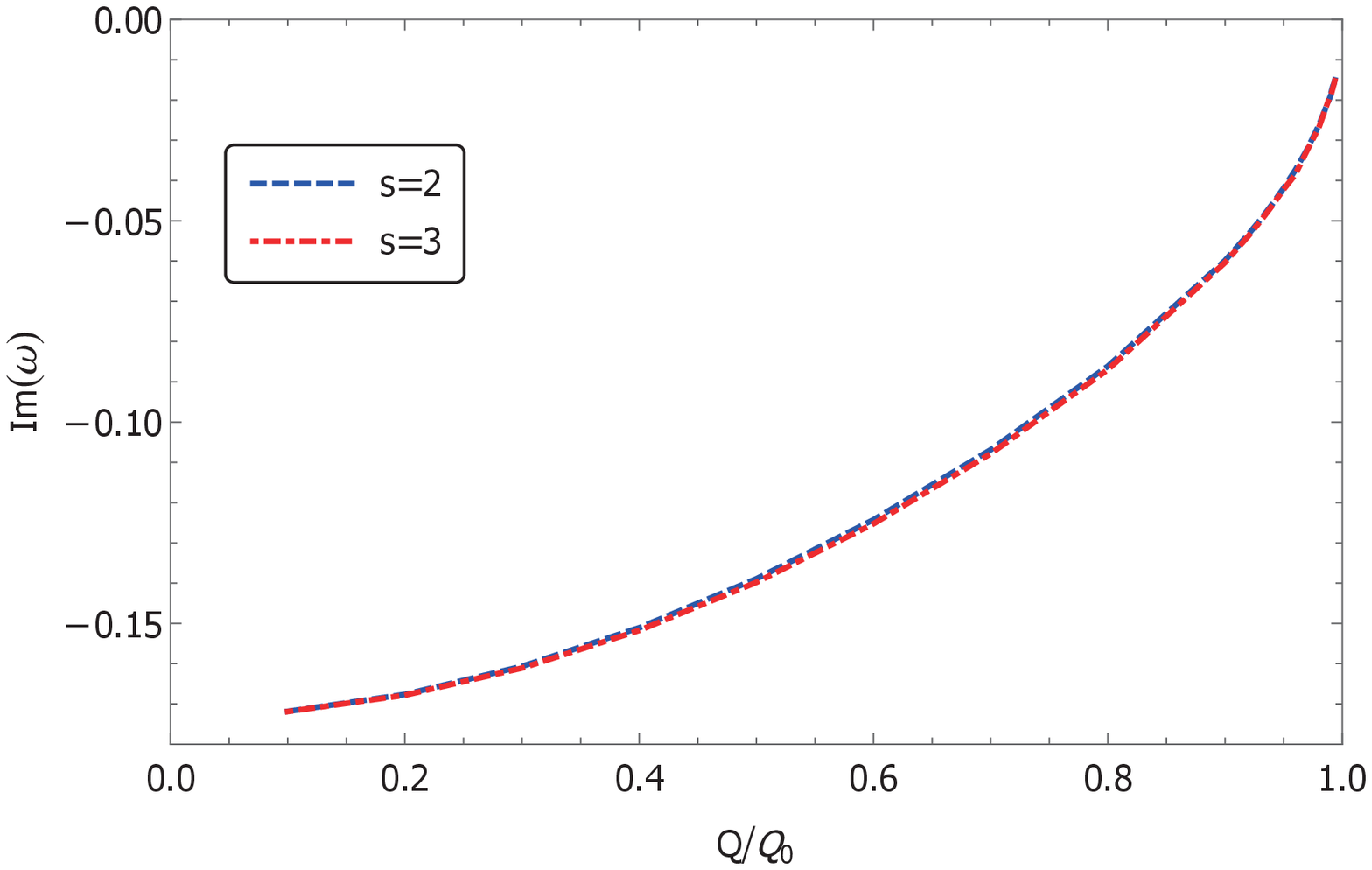}
\caption{(Color Online) The calculated quasinormal frequencies for massless Dirac field as functions of the charge for the extremal black holes regarding the Nariai solution.
The calculations are carried out for $s=2$ and $3$ with $w=-1$.}\label{Fig2}
\end{figure*}

Next, we analyze the quasinormal frequencies of the second type of extremal black hole metric, discussed in the previous section regarding Eqs.~(\ref{N35}) and (\ref{N38}).
As shown in Fig.~\ref{Fig2}, as the charge of black hole approaches the corresponding extremal value $Q_0$, both the real and imaginary parts of the quasinormal frequency go to zero.
The asymptotical behavior can be understood as follows.
Due to physical constraints, the wave function must satisfy the following boundary conditions, namely, it is asymptotically an outgoing wave at the cosmological horizon and an ingoing wave at the event horizon.
As ${r_h}$ approaches ${r_c}$, the waveform of the outgoing wave shall coincide with that of the ingoing wave.
Subsequently, it implies $\omega=0$.
In other words, the quasinormal frequency should be vanishing in the case of the extremal black hole, as observed in numerical calculations.

\begin{figure*}[htbp]
\centering
\includegraphics[scale=0.33]{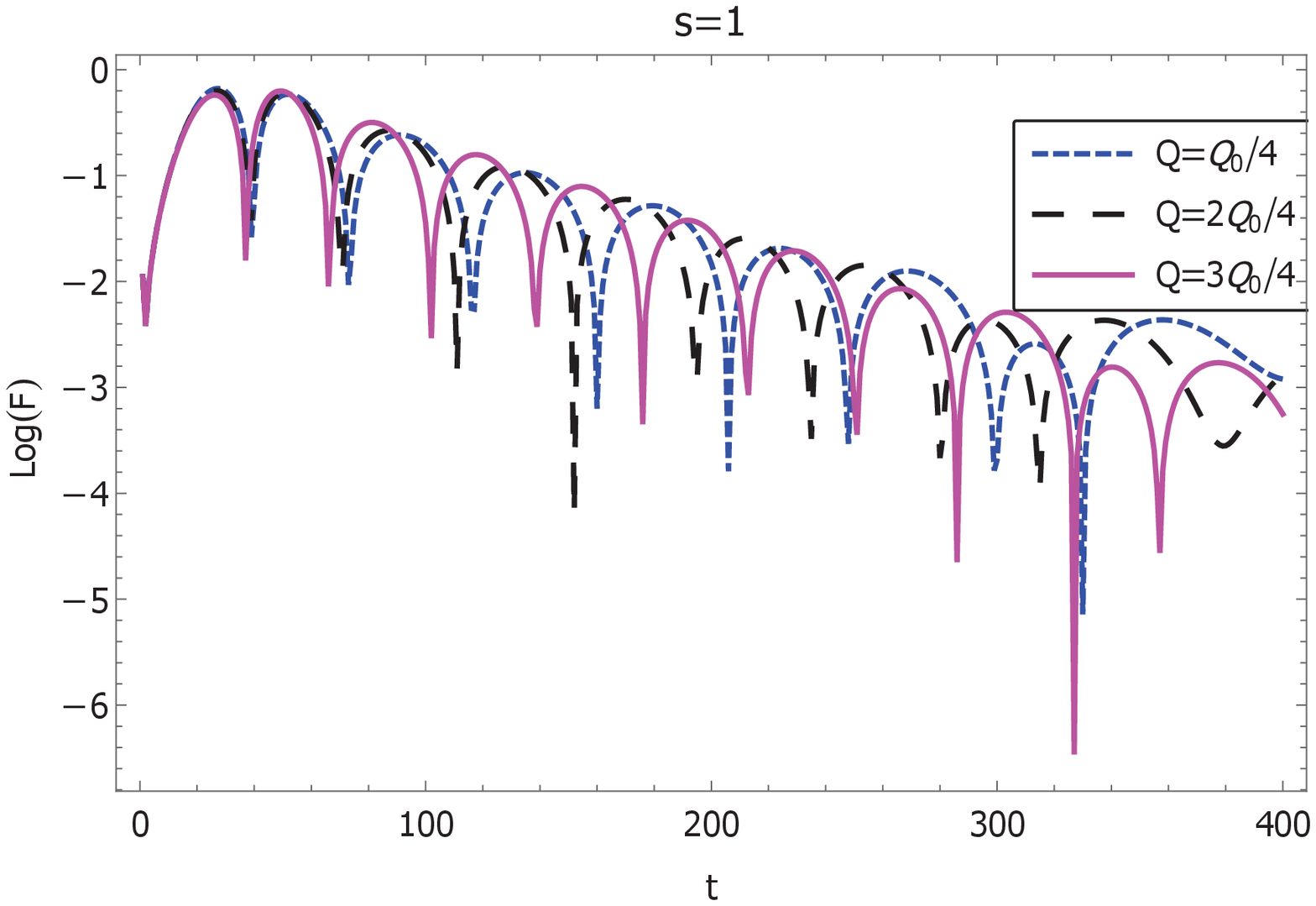}
\includegraphics[scale=0.33]{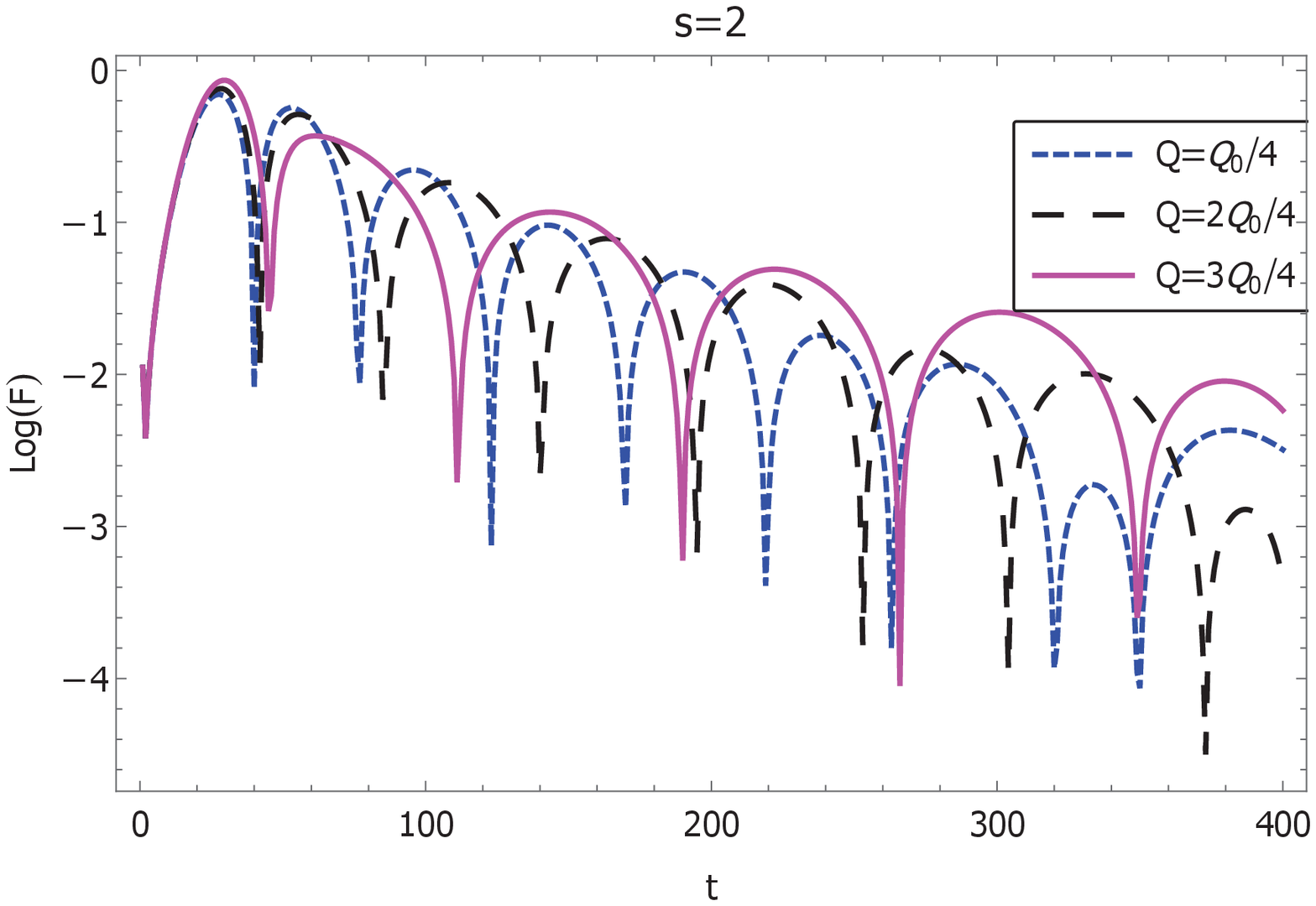}
\includegraphics[scale=0.33]{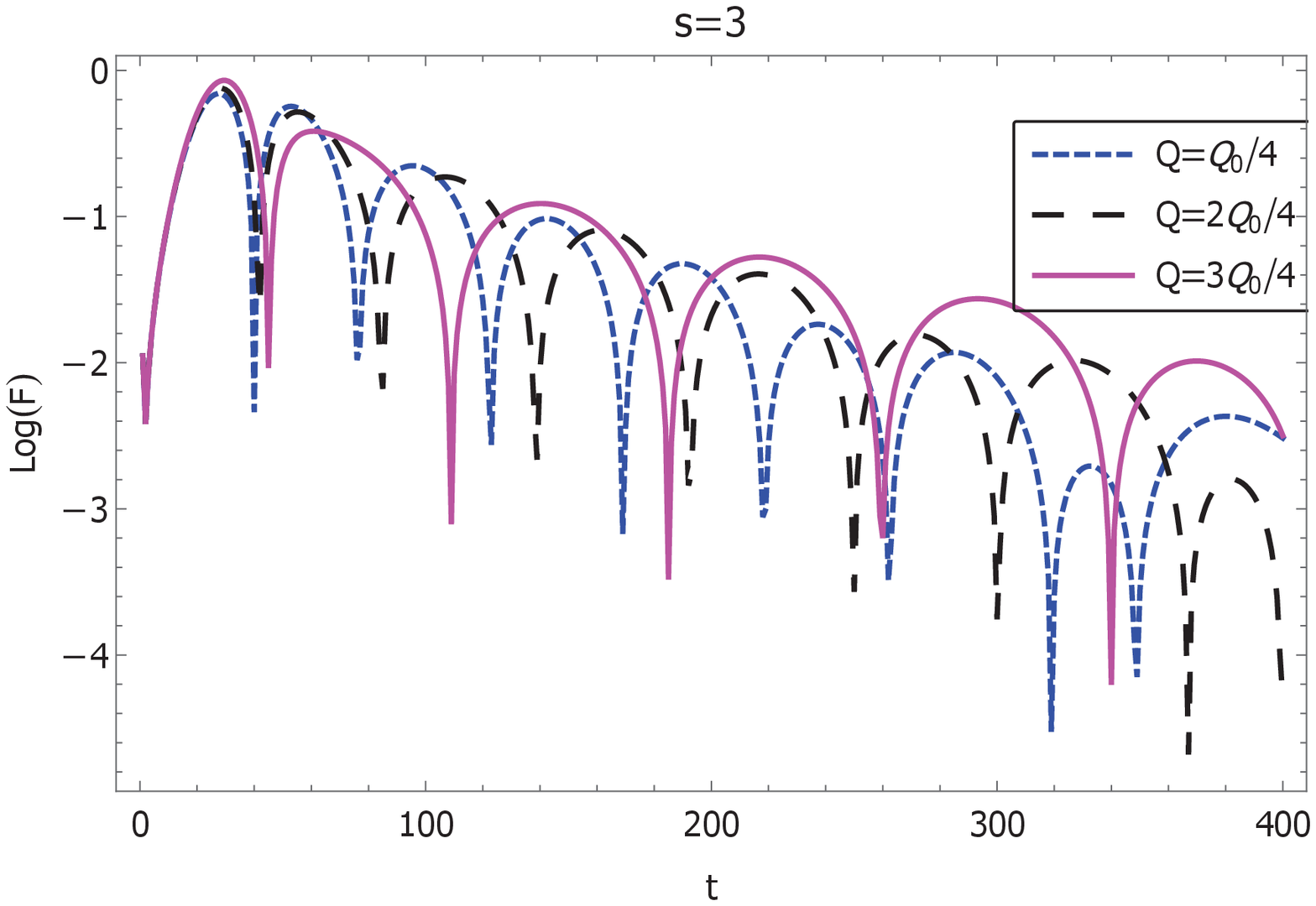}
\caption{(Color Online) The temporal evolution of the massless Dirac perturbations for different black hole spacetimes.
The calculations are carried out for different $s$ and $Q$ with given $\left| \kappa \right| = 2$.}\label{Fig5}
\end{figure*}

In what follows, we present the temporal evolution of the initial perturbations of a massless Dirac field by using the finite difference method.
It can be seen from Fig.~\ref{Fig5} initial outburst of the waves stabilizes in time and subsequently turns into quasinormal oscillations.
The time scale for the system to reach quasinormal oscillations depend on the charge of the black hole.
For the metrics with $s=1$, the quasinormal oscillations tend to stabilize faster as the charge increases.
On the contrary, for the metrics with $s=2$ and $3$, the first stage of initial outburst persist longer as the charge becomes larger.

Moreover, it might be interesting to compare the above results against other forms of perturbations.
It is observed that the magnitudes of the real parts of the quasinormal frequencies are the largest for the Dirac perturbations, while those of the imaginary parts are most significant for the scalar perturbations.  
Besides, the magnitudes for the electromagnetic perturbations are found to be less significant, and furthermore, those for the gravitational perturbations are often the least.
These results are not shown explicitly here due to the scope of the present study. 
Nonetheless, we note that they are largely consistent with the existing findings in the literature~\cite{ctp-qnm-rastall}.

Last but not least, we investigate the asymptotic properties of the late-time tails of quasinormal modes in the present model.
As discussed above, by appropriate choice of parameters, the black hole metric in Rastall theory for asymptotically flat spacetimes are equivalent to those in Einstein gravity for asymptotically de Sitter spacetimes.
However, for Einstein's gravity, it is known that the late-time tails in asymptotical spacetime follow a power-law form while those for asymptotically de Sitter spacetimes decay exponentially.
Therefore, in what follows, we carry out explicitly calculations by focusing on the features of the last stage of quasinormal modes.
In particular, we also study a specific scenario where the asymptotical behavior of the spacetimes again falls back to that of flat spacetimes.
The latter is achieved by choosing the parameters so that the last term on the r.h.s. of Eq.~\eqref{N10} becomes irrelevant.

\begin{figure*}[htbp]
\centering
\includegraphics[scale=0.5]{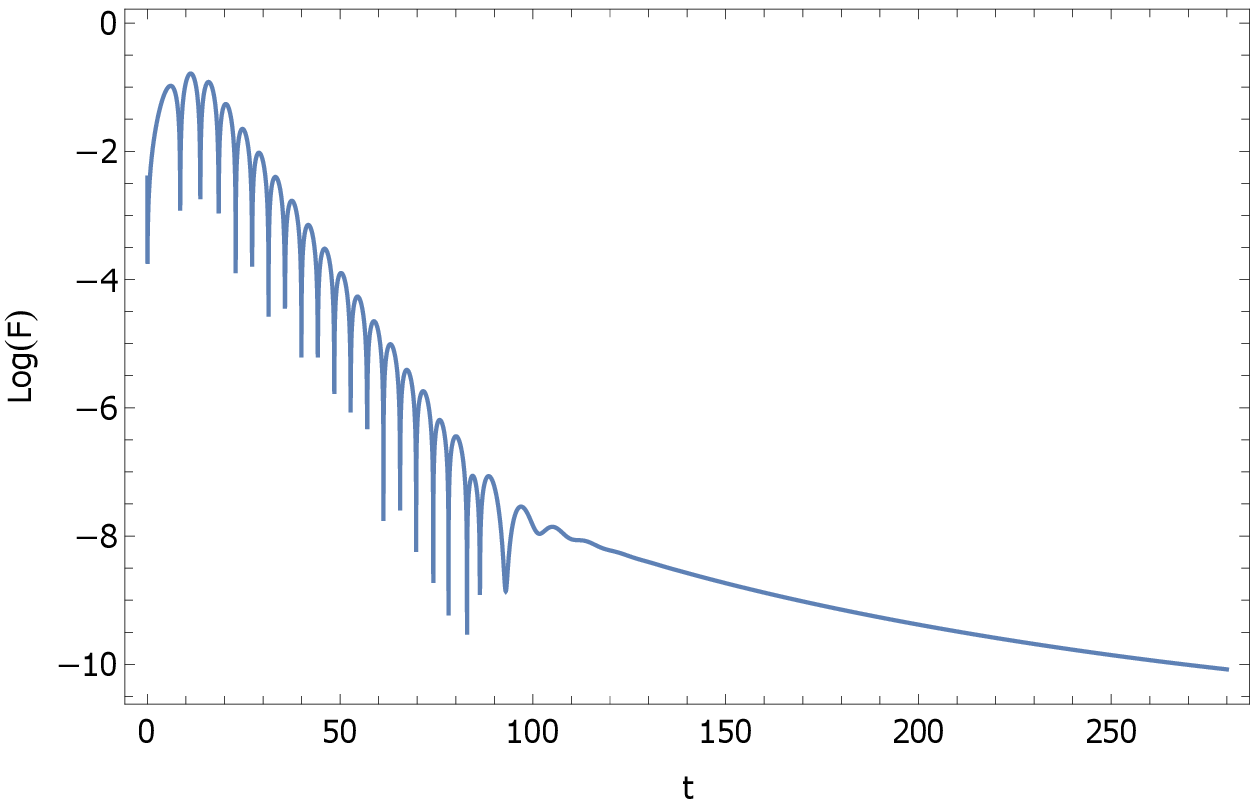}
\includegraphics[scale=0.55]{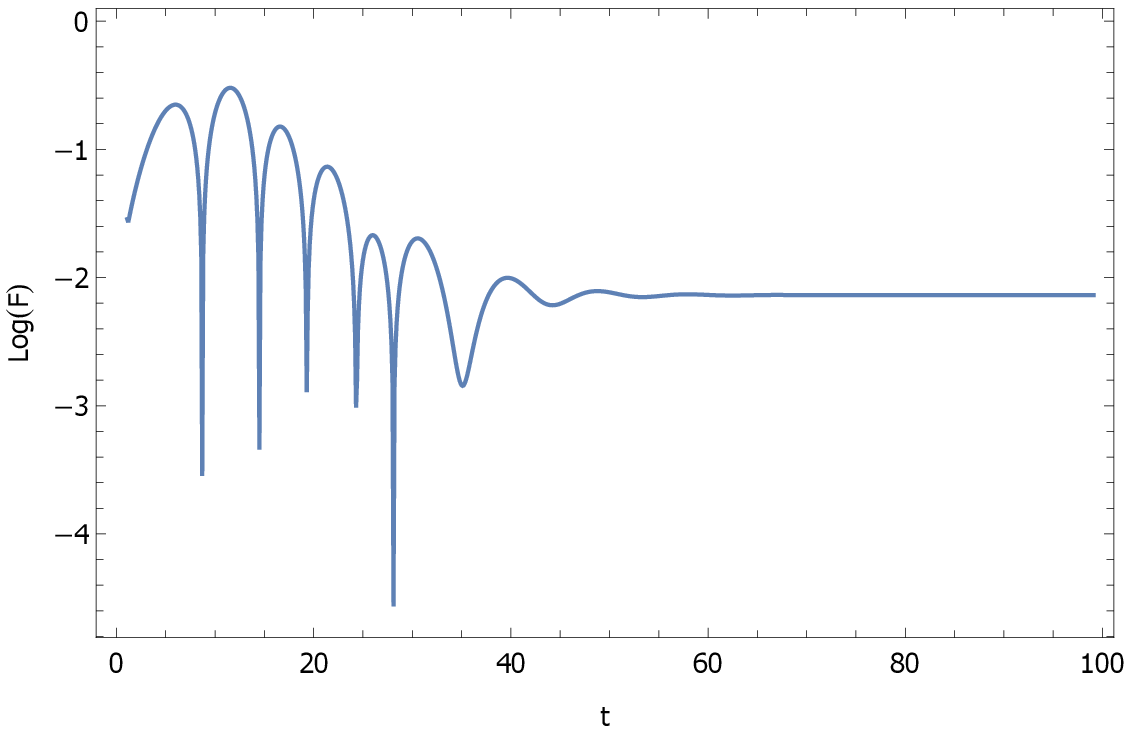}
\caption{(Color Online) The temporal evolution of the massless Dirac perturbations for different black hole spacetimes by using the specific metrics discussed in the text.
The calculations are carried out to demonstrate the specific features of the late-time tail of the quasinormal modes.
}\label{Fig6}
\end{figure*}

We first consider the case $Q = 0$ and $\omega=-1$ in Eq.~\eqref{N11}, which implies
\begin{equation}\label{N47}
{f}(r) = 1 - \frac{{2M}}{r} - \frac{{{r^2}}}{9}.
\end{equation}
This corresponds a black hole metric in an asymptotically flat spacetime for Rastall gravity, which is equivalent to that in an asymptotically de Sitter spacetime.
By solving the master equation Eq.~\eqref{N45}, the temporal evolution is evaluated and shown in Fig.~\ref{Fig6}.
It is observed that the late-time tail decays exponentially at late times.
By carrying out a $\chi^2$ fitting, it is found that its function form satisfies ${e^{ - \alpha t}}$ with $\alpha=0.02503  \pm 0.00019 $.
In other words, the late-time tail is indeed similar to the asymptotical behavior of a massless Dirac perturbation in a Schwarzschild-de Sitter black hole metric for Einstein gravity.

On the other hand, one assumes $Q = 0$ and $\omega=-\frac{1}{3}$ in Eq.~\eqref{N11}, which implies
\begin{equation}\label{N48}
{f}(r) = 1 - \frac{{2M}}{r} - \frac{1}{{9{r^4}}}.
\end{equation}
It is not difficult to show that the related metric is asymptotically flat.
By performing similar numerical calculations, one finds that the perturbations decay according to the power-law at the late-time, as shown in Fig.~\ref{Fig6}.
The decay rate of the tail is found to be $t^{ - \beta}$ with $\beta=1.76676 \pm 0.00031 $.
Again, this is in agreement with the case of a Schwarzschild black hole metric in asymptotically flat spacetime for Einstein gravity.

Therefore, one concludes that the asymptotical behavior of the late-time tail of quasinormal modes in Rastall theory follows their counterpart in Einstein gravity, in accordance with the asymptotical properties of the spacetimes~\cite{Ching:1995tj, Brady:1999wd}.

\section{VI. Concluding remarks}\label{section6}

To summarize, in this work, the quasinormal modes of the massless Dirac field is investigated for charged black holes in Rastall gravity.
We explored the recently established spherically, symmetric black hole solutions.
The latter is characterized by the linear or nonlinear power-Maxwell field, surrounded by the quintessence fluid.
We have carried out numerical calculations by employing the WKB approximations up to the thirteenth order, as well as the matrix method.
The results obtained by the WKB approximations and matrix method are found to be consistent.
Also, the dependence on the model parameters is investigated regarding their effects on the quasinormal frequencies.
Besides, we explore the properties of a second type of extremal black holes regarding the Nariai solution, as well as the related quasinormal modes.
Moreover, the finite difference method is also utilized to study the temporal evolution of the small initial perturbations.
In particular, one finds that the asymptotical behavior of the late-time tails of quasinormal modes in Rastall theory is dictated by their counterparts in Einstein gravity, reflecting the asymptotical properties of the spacetimes.
We plan to further generalize the present study to the case of rotating black holes in high-dimensional Rastall spacetimes.

\section*{Acknowledgements}

We gratefully acknowledge the financial support
from National Natural Science Foundation of China (NNSFC) under contract No. 11805166, 11805074, 11925503, and 91636221,
Post-doctoral Science Foundation of China under contract No. 2018T110750,
and Brazilian funding agencies Funda\c{c}\~ao de Amparo \`a Pesquisa do Estado de S\~ao Paulo (FAPESP),
Conselho Nacional de Desenvolvimento Cient\'{\i}fico e Tecnol\'ogico (CNPq), Coordena\c{c}\~ao de Aperfei\c{c}oamento de Pessoal de N\'ivel Superior (CAPES).

\bibliographystyle{h-physrev}
\bibliography{references_shao}

\end{document}